\documentclass[12pt,draftclsnofoot,onecolumn]{IEEEtran}

\usepackage{amsmath,amsfonts,amssymb,bm}
\usepackage[tight,footnotesize]{subfigure}
\usepackage{graphicx,stfloats}
\usepackage{algorithm}
\usepackage{algorithmic}
\usepackage{cite,url}
\usepackage{cases}
\usepackage{extarrows}
\usepackage{theorem} % for \theorembodyfont{\rmfamily}
\usepackage{paralist}
\usepackage{psfrag}
\usepackage{auto-pst-pdf}
\usepackage{subfigure}
\usepackage{bm}
\usepackage{float}

\allowdisplaybreaks[4]  % 全局强制公式分页
\theorembodyfont{\rmfamily}

\newtheorem{Theorem}{Theorem}

\newtheorem{Lemma}{Lemma}

\newcommand{\Lset}{\ensuremath{\mathcal{L}}}
\newcommand{\Cset}{\ensuremath{\mathbb{C}}}
\newcommand{\Eset}{\ensuremath{\mathbb{E}}}
\newcommand{\Rset}{\ensuremath{\mathbb{R}}}

\makeatletter

\newcommand{\Rmnum}[1]{\expandafter\@slowromancap\romannumeral #1@}
\makeatother

\begin{document}
\vspace{-2ex}
\title{\LARGE Massive MIMO with Multi-cell MMSE Processing: Exploiting All Pilots for Interference Suppression}
\author{{\normalsize Xueru~Li$^\star$,~\IEEEmembership{}
        Emil~Bj{\"o}rnson$^{\dagger}$,~\IEEEmembership{\normalsize Member,~IEEE,}
        Erik~G.~Larsson$^{\dagger}$,~\IEEEmembership{\normalsize Senior~Member,~IEEE,}
        Shidong~Zhou$^\star$,~\IEEEmembership{\normalsize Member,~IEEE,}
        and~Jing~Wang$^\star$,~\IEEEmembership{\normalsize Member,~IEEE,}}\vspace{-3ex}% <-this % stops a space
\thanks{ $^\star$X.~Li, S.~Zhou and J.~Wang are with the Department of Electronic Engineering, Research Institute of Information Technology, Tsinghua University, Beijing 100084, China. E-mail: xueruli1206@163.com, \{zhousd, wangj\}@mail.tsinghua.edu.cn.}
\thanks{ $^{\dagger}$E.~Bj{\"o}rnson and E.~G.~Larsson are with the Department of Electrical Engineering (ISY), Link{\"o}ping University, SE-58183 Link{\"o}ping, Sweden. E-mail: \{erik.g.larsson, emil.bjornson\}@liu.se.}
\thanks{ This work was performed when X. Li was a visiting Ph.D. student at Link\"oping University. The work is supported by National Basic Research Program (2012CB316000), National Natural Science Foundation of China (61201192), National High Technology Research Development Program of China (2014AA01A703), National S\&T Major Project (2014ZX03003003-002), Tsinghua-Qualcomm Joint Research Program, Keysight Technologies, Inc., ELLIIT, the CENIIT project 15.01 and FP7-MAMMOET.}\vspace{-5ex}}

\maketitle
\begin{abstract}
In this paper, a new state-of-the-art multi-cell MMSE scheme is proposed for massive MIMO networks, which includes an uplink MMSE detector and a downlink MMSE precoder. The main novelty is that it exploits all available pilots for interference suppression. Specifically, let $K$ and $B$ denote the number of users per cell and the number of orthogonal pilot sequences in the network, respectively, where $\beta = B/K$ is the pilot reuse factor. Then our multi-cell MMSE scheme utilizes all $B$ channel directions, that can be estimated locally at each base station, to actively suppress both intra-cell and inter-cell interference. The proposed scheme is particularly practical and general, since power control for the pilot and payload, imperfect channel estimation and arbitrary pilot allocation are all accounted for. Simulations show that significant spectral efficiency (SE) gains are obtained over the single-cell MMSE scheme and the multi-cell ZF, particularly for large $\beta$ and/or $K$. Furthermore, large-scale approximations of the uplink and downlink SINRs are derived, which are asymptotically tight in the large-system limit. The approximations are easy to compute and very accurate even for small system dimensions. Using these SINR approximations, a low-complexity power control algorithm is also proposed to maximize the sum SE.
\end{abstract}

\begin{IEEEkeywords}
Massive MIMO, multi-cell MMSE processing, large-scale SINR approximations, power control.
\end{IEEEkeywords}

\section{Introduction} \label{intro}
Multi-user multiple-input-multiple-output (MU-MIMO) communication has drawn considerable interest in recent years. %It offers big advantages over conventional point-to-point MIMO communication.
By scheduling multiple users to share the spatial channel simultaneously, the spatial degrees of freedom offered by multiple antennas can be exploited to focus signals on intended receivers, reduce interference, and thereby increase the system data rate~\cite{Caire2010,Caire2003,Wei2004,Gesbert2007,Stankovic2008,Yoo2006}. These features make MU-MIMO incorporated into recent and evolving wireless standards like 4G long-term evolution (LTE) and LTE-Advanced~\cite{Larsson14}.

Massive MU-MIMO, or very large MU-MIMO, is an emerging technology that scales up MU-MIMO by orders of magnitude~\cite{Marzetta10,Rusek13}. The idea is to employ an array comprising say a hundred, or more, antennas at the base station (BS) and serve tens of users simultaneously per cell. Compared to the contemporary cellular systems, the system SE can be drastically increased without consuming extra bandwidth~\cite{Marzetta10, Rusek13, Larsson14}. Uplink and downlink transmit power can also be reduced by an order of magnitude since the phase-coherent processing provides a comparable array gain~\cite{Hien13}. In the limit of an infinite number of antennas, intra-cell interference and uncorrelated noise can be averaged out by using simple coherent precoders and detectors, and the only performance limitation is pilot contamination and the distortion noise from hardware impairments~\cite{Marzetta10,bjornson2013massive}. Furthermore, in time division duplex (TDD) mode, the channel training overhead scales linearly with the number of users, instead of the number of BS antennas, which allows for adding antennas elements without affecting the training overhead~\cite{Marzetta2006}. These features make massive MIMO one of the key technologies for the next generation wireless communication systems.

In the uplink reception and downlink transmission, the most common linear processing schemes are matched filtering (MF), zero forcing (ZF) and minimum mean square error (MMSE).\footnotemark{ }Let $B$ denote the number of orthogonal pilot sequences that are available in the network, and $K$ denote the number of users in each cell. We can then define $\beta = B/K \ge 1$ as the pilot reuse factor, since only $1/\beta$ of the cells use the same set of pilots. In conventional massive MIMO systems, the BS first listens to the uplink pilot signalling from its own cell, estimates the $K$ intra-cell channels and then constructs its transceiver processing based on the channel estimates to mitigate the intra-cell interference~\cite{Hoydis2013,Guo2014,Krishnan2014,Hien2012}. However, parts of the inter-cell interference can also be suppressed when $\beta > 1$. If the BS is aware of all pilot sequences, then it can locally estimate $B$ channel directions by listening to the pilot signalling from all cells instead of only from its own cell. Since its $K$ users only occupy $K$ out of the $B$ channel directions, the BS is able to select its user-specific detectors in the uplink to suppress interference from other cells, and design precoders in the downlink to mitigate interference leakage to other cells. Based on similar observations, some multi-cell detection and precoding schemes have been proposed in~\cite{bjornson2014massive,Hien2012,Jose2011,Guo2013}. In~\cite{bjornson2014massive}, a multi-cell ZF detector (referred to as full-pilot ZF detector in~\cite{bjornson2014massive}) is proposed, which exploits and orthogonalizes all available directions to mitigate parts of the inter-cell interference. It achieves a higher SE than the conventional ZF when the interfering users are near to the edges of the surrounding cells. In general cellular networks, however, the gain is less obvious, partly due to the loss in array gain of $B$ in multi-cell ZF, instead of $K$ as with conventional ZF. Uplink multi-cell MMSE detectors are proposed in~\cite{Hien2012} and~\cite{Guo2013}, but the former is limited to $\beta =1$ and equal power allocation, and the latter is based on the unrealistic assumption that perfect channel state information (CSI) is known at BS. The multi-cell MMSE precoder proposed in~\cite{Jose2011} brings a notable gain over single-cell processings. However, like~\cite{Hien2012}, this scheme does not account for arbitrary pilot allocation which, as shown in~\cite{bjornson2014massive}, is an important way to suppress pilot contamination and achieve high system SE in massive MIMO deployments. Moreover, no closed-form performance expressions are provided in~\cite{Jose2011}.
\footnotetext{A special case of the downlink MMSE precoder is the regularized ZF (RZF) precoder, which is obtained when all the users in a cell have equal pathlosses~\cite{bjornson2014optimal}. Since this is generally not the case in cellular networks, RZF provides lower performance than the MMSE precoder and is not considered in this paper.}

In this paper, a new state-of-the-art multi-cell MMSE transceiver scheme is proposed, which includes an uplink MMSE detector and a downlink transmit MMSE precoder. The novelty of the multi-cell MMSE scheme is that all $B$ pilots are exploited at each BS to actively suppress both intra-cell and inter-cell interference. Power control for the pilot and payload, imperfect channel estimation and arbitrary pilot allocation are all accounted for in our scheme. Numerical results show that significant SE gains can be obtained by the proposed scheme over conventional single-cell schemes and the multi-cell ZF from~\cite{bjornson2014massive}, and the gains become more significant as $\beta$ and/or $K$ increase. Furthermore, large-scale approximations of the uplink and downlink SINRs are derived for the proposed multi-cell MMSE scheme, which are asymptotically tight in the large-system limit. The approximations are easy to compute since they only depend on large-scale fading, power control and pilot allocation, and shown to be very accurate even for small system dimensions. Based on the SINR approximations, a low complexity iterative power control algorithm for sum SE maximization is proposed for the multi-cell MMSE scheme. Compared to the equal power allocation policy, our proposed algorithm significantly improves the system sum SE and also provides good user fairness.

The paper is organized as follows: In Section~\ref{system model}, we describe the system model and the construction of the multi-cell MMSE transceiver. Large-scale approximations of the uplink and downlink SINRs are derived in Section~\ref{asymptotic analysis}. Based on the SINR approximations, a low complexity iterative power control algorithm is proposed in Section~\ref{sec:powercontrol}. Simulation results are provided in Section~\ref{simulation} before we conclude the paper in Section~\ref{conclusions}. All proofs are deferred to the appendix.

\textit{Notations}: Boldface lower and upper case symbols represent vectors and matrices, respectively. The trace, transpose, conjugate, Hermitian transpose and matrix inverse operators are denoted by $\rm{tr}(\cdot )$, $(\cdot )^T$, $(\cdot )^{*}$, $(\cdot )^H$ and $(\cdot )^{-1}$, respectively.

\section{System Model and Transceiver Design} \label{system model}
We consider a synchronous massive MIMO cellular network with multiple cells. Each cell is assigned with an index in the cell set $\Lset$, and the cardinality $| \Lset |$ is the number of cells. The BS in each cell is equipped with an antenna array of $M$ antennas and serves $K$ single-antenna users within each coherence block. Assume that this time-frequency block consists of $T_c$ seconds and $W_c$ Hz, such that $T_c$ is smaller than the coherence time of all users and $W_c$ is smaller than the coherence bandwidth of all users. This leaves room for $S=T_c \times W_c$ transmission symbols per block, and the channels of all users remain constant within each block. Let ${{\bf{h}}_{jlk}}$ denote the channel response from user $k$ in cell $l$ to BS $j$ within a block, and assume that it is a realization from a zero-mean circularly symmetric complex Gaussian distribution:
\begin{equation} \label{channel model}
{{\bf{h}}_{jlk}} \sim {\cal{CN}}\left( {{\bf 0},{d_j}\left( {{{\bf{z}}_{lk}}} \right){{\bf{I}}_M}} \right).
\end{equation}
The vector ${\bf{z}}_{lk} \in \Rset ^2$ is the geographical position of user $k$ in cell $l$ and ${{d_j}( {{{\bf{z}}}})}$ is an arbitrary function that accounts for the channel attenuation (e.g., path loss and shadowing) between BS $j$ and any user position $\bf{z}$. Since the user position changes relatively slowly, ${{d_j}( {{{\bf{z}}_{lk}}})}$ is assumed to be known at BS $j$ for all $l$ and all $k$.

We consider a TDD protocol in this paper. where the downlink channels are estimated by uplink pilot signaling by exploiting channel reciprocity. In TDD mode, each transmission block is divided into two phases: $\left.1\right)$ uplink channel estimation phase, where each BS estimates the CSI from uplink pilot signalling which occupies $B$ out of $S$ symbols in each block; $\left.2\right)$ uplink and downlink payload data transmission phase, where each BS processes the received uplink signal and the to-be-transmitted downlink signals using the estimated CSI. Let $\zeta^{\rm{ul}}$ and $\zeta^{\rm{dl}}$ denote the fixed fractions allocated for uplink and downlink payload data transmission, respectively. These fractions can be selected arbitrarily under the conditions that $\zeta^{\rm{ul}}+\zeta^{\rm{dl}}=1$ and that $\zeta^{\rm{ul}}(S-B)$ and $\zeta^{\rm{dl}}(S-B)$ are positive integers. In what follows, the uplink channel estimation is first discussed to lay a foundation for the transceiver design.
\vspace{-1ex}
%\section{Uplink Channel Estimation} \label{channel estimation}
\subsection{Uplink Channel Estimation} \label{channel estimation}
In the uplink channel estimation phase, the collective received signal at BS $j$ is denoted as ${{\bf{Y}}_j} \in {\Cset^{M \times B}}$ where $B$ is the length of the pilot sequences (it also equals to the number of orthogonal pilot sequences available in the network). Then ${\bf Y}_j$ can be expressed as
\begin{equation} \label{estimationsignal}
{{\bf{Y}}_j} = \sum\limits_{l \in {\cal L}} {\sum\limits_{k = 1}^K {\sqrt {{p_{lk}}} {{\bf{h}}_{jlk}}{\bf{v}}_{{i_{lk}}}^T} }  + {{\bf{N}}_j},
\end{equation}
where ${\bf{h}}_{jlk}$ is the channel response defined in~(\ref{channel model}), $p_{lk}\geq 0$ is the power control coefficient for the pilot of user $k$ in cell $l$, and ${{\bf{N}}_j}\in\Cset^{M \times B}$ contains independent and identically distributed (i.i.d.) elements that follow ${\cal {CN}}(0,\sigma^2)$. We assume that all pilot sequences originate from a predefined orthogonal pilot book, defined as ${\cal{V}}= \{ {{{\bf{v}}_1},\ldots,{{\bf{v}}_B}}\}$, where
\begin{equation}
{\bf{v}}_{{b_1}}^H{{\bf{v}}_{{b_2}}} = \left\{ \begin{array}{ll}
B,  & {b_1} = {b_2},\\
0,  & {b_1} \ne {b_2},
\end{array} \right.
\end{equation}
and let ${i_{lk}} \in \{ 1,\ldots,B\}$ denote the index of the pilot sequence used by user $k$ in cell $l$. Arbitrary pilot reuse is supported in our work by denoting the relation between $B$ and $K$ by $B = \beta K$, where $\beta \ge 1$ is called the pilot reuse factor. If the pilots are allocated wisely in the network, a larger $\beta$ brings a lower level of interference during pilot transmission, known as pilot contamination. %, since a smaller fraction of cells use the same pilot sequences as the target cell.

Based on the received signal in~(\ref{estimationsignal}), the MMSE estimate of the uplink channel ${{\bf{h}}}_{jlk}$ is~\cite{bjornson2014massive}
\begin{equation} \label{estimation}
{\hat{\bf h}}_{jlk} = \sqrt{p_{lk}} d_j\left( {\bf z}_{lk}\right){\bf Y}_j \left({\bf \Psi}_j^{*}\right)^{-1} {\bf v}_{i_{lk}}^{*},
\end{equation}
where ${{\bf{\Psi }}_j}$ is the covariance matrix of the vectorized received signal ${\rm{vec}}({\bf Y}_j)$ and is given by
\begin{equation}
{{\bf{\Psi }}_j} =\Eset\left\{ {\rm{vec}}\left({\bf Y}_j\right){\rm{vec}}\left({\bf Y}_j\right)^H \right\}= \sum\limits_{\ell \in {\cal L}} {\sum\limits_{m = 1}^K {p_{{\ell m}}} d_j\left( {\bf z}_{\ell m}\right){\bf{v}}_{{i_{\ell m}}}{\bf{v}}_{{i_{\ell m}}}^H} + \sigma ^2{{\bf{I}}_B}.
\end{equation}
According to the orthogonality principle of MMSE estimation, the covariance matrix of the estimation error ${\tilde {\bf h}}_{jlk} ={{\bf{h}}_{jlk} - {\hat{\bf h}}_{jlk}}$ is given by
\begin{eqnarray}\label{errormatrix}
{{\bf{C}}_{jlk}} &=& {\Eset}\left\{  {{\tilde {\bf{h}}}_{jlk} { {{\tilde {\bf{h}}}_{jlk}^H}}} \right\}
= d_j \left( {{{\bf{z}}_{lk}}} \right)  \left( {1 - p_{lk} d_j \left( {{{\bf{z}}_{lk}}} \right){\bf{v}}_{{i_{lk}}}^H{\bf{\Psi }}_j^{ - 1}{{\bf{v}}_{{i_{lk}}}}} \right){{\bf{I}}_M}.
\end{eqnarray}
By utilizing that
\begin{equation} \label{temp}
{\bf{v}}_{{i_{lk}}}^H{\bf{\Psi }}_j^{ - 1} = \underbrace {\frac{1}{{\sum\nolimits_{\ell  \in {\cal L}} {\sum\nolimits_{m = 1}^K { p_{\ell m}{{{d_j}\left( {{{\bf{z}}_{\ell m}}} \right)}}{\bf{v}}_{{i_{lk}}}^H{{\bf{v}}_{{i_{\ell m}}}}} }  + \sigma ^2}}}_{{\alpha _{ji_{lk}}}}{\bf{v}}_{{i_{lk}}}^H = {\alpha _{ji_{lk}}}{\bf{v}}_{{i_{lk}}}^H,
\end{equation}
where $\alpha_{ji_{lk}}$ is a scalar, the estimation error covariance matrix in~(\ref{errormatrix}) can be expressed as
\begin{equation} \label{errormatrix2}
{{\bf{C}}_{jlk}} = d_j \left( {{{\bf{z}}_{lk}}} \right)  \left( 1 - p_{lk} d_j \left( {{{\bf{z}}_{lk}}}  \right) {\alpha _{ji_{lk}}}B\right){{\bf{I}}_M}.
\end{equation}

As pointed out in~\cite{bjornson2014massive}, the part ${\bf Y}_j({\bf \Psi}_j^{*})^{-1}{\bf v}_{i_{lk}}^{*}$ of the MMSE channel estimate in~(\ref{estimation}) depends only on which pilot sequence that user $k$ in cell $l$ uses. Consequently, users who use the same pilot sequence have parallel estimated channels at each BS, while only the amplitudes are different in the estimates. To show this explicitly, define the $M \times B$ matrix
\begin{eqnarray}
{{{\hat {\bf H}}}_{\mathcal{V},j}}=\left[ {\hat{\bf h}}_{{\cal V},j1},...,{\hat{\bf h}}_{{\cal V},jB} \right] = {\bf Y}_j \left({\bf \Psi}_j^{*}\right)^{-1} \left[{\bf v}_1^{*},...,{\bf v}_B^{*}\right],
\end{eqnarray}
which allows the channel estimate in~(\ref{estimation}) to be reformulated as
\begin{equation} \label{estimation2}
{\hat {\bf h}}_{jlk} = \sqrt{p_{lk}} d_j \left( {{{\bf{z}}_{lk}}} \right){{{\hat {\bf H}}}_{{\cal V},j}}{{\bf{e}}_{{i_{lk}}}},
\end{equation}
where ${{\bf{e}}_i}$ denotes the $i$th column of the identity matrix ${\bf I}_B$. The property that users with the same pilot have parallel estimated channels is utilized to derive new SE expressions in the sequel.

Notice that the estimated channel ${\hat{\bf h}}_{jlk}$ is also a zero-mean complex Gaussian vector, with its covariance matrix ${{\bf{\Phi }}_{jlk}}\in \mathbb{C}^{M \times M}$ being
\begin{equation} \label{covariance}
{{\mathbf{\Phi }}_{jlk}} = d_j \left( {{{\bf{z}}_{lk}}} \right){\bf I}_M - {\bf C}_{jlk} = {p_{lk}} d_j^2\left( {{{\bf{z}}_{lk}}}\right){\alpha _{ji_{lk}}}B {\bf I}_M.
\end{equation}
Define the covariance matrix of ${\hat {\bf h}}_{{\cal V},ji}$ as ${ \tilde {\bf \Phi}}_{{\cal{V}},ji}$. Then according to~(\ref{estimation2}) and~(\ref{covariance}), ${\tilde {\bf \Phi}}_{{\cal V},j{i}} = \alpha_{ji}B {\bf I}_M$.

\subsection{Uplink Multi-cell MMSE detector} \label{mmse detector}
After the uplink channel estimation, during the uplink payload data transmission phase, the received signal ${{\bf{y}}_j} \in \Cset^{M \times 1}$ at BS $j$ is
\begin{equation} \label{signalmodel}
{{\bf{y}}_j} = \sum\limits_{l \in \cal L} {\sum\limits_{k = 1}^K {\sqrt {{\tau_{lk}}} {{\bf{h}}_{jlk}}{x_{lk}}} }  + {{\bf{n}}_j},
\end{equation}
where $\tau_{lk}$ is the transmit power of the payload data from user $k$ in cell $l$,  $x_{lk} \sim {\cal{CN}}(0,1)$ is the transmitted signal from a Gaussian codebook, and ${\bf n}_j \sim {\cal{CN}}({\bf 0},\sigma^2 {\bf I}_M )$ is additive white Gaussian noise (AWGN). Different symbols are used for pilot power and payload power to allow for different power control policies for them. Denote the linear detector used by BS $j$ for an arbitrary user $k$ in its cell as ${\bf{g}}_{jk}$, then the detected signal ${\hat x}_{jk}$ is
\begin{equation} \label{x_detected}
{\hat x}_{jk} ={\bf{g}}_{jk}^H {\bf y}_j = \sqrt{\tau_{jk}} {\bf g}_{jk}^H{\bf{h}}_{jjk} x_{jk} + {\bf g}_{jk}^H \sum\limits_{\left(l,m\right) \ne \left(j,k \right)}\sqrt{\tau_{lm}} {\bf h}_{jlm}x_{lm} + {\bf g}_{jk}^H  {\bf n}_j.
\end{equation}
By using~(\ref{x_detected}), the following achievable ergodic SE can be achieved for this user~\cite{Hoydis2013}
\begin{equation} \label{rate_ul}
R_{jk}^{\rm{ul}} = \zeta^{\rm{ul}} \left(1-\frac{B}{S} \right) \Eset_{\left\{{\hat {\bf h}}_{\left(j \right)}\right\}} \left\{ \log_2\left( 1+ \eta_{jk}^{\rm{ul}}\right) \right\},
\end{equation}
where $\Eset_{\{{\hat {\bf h}}_{(j )}\}}$ denotes the expectation with respect to all the channel estimates obtained at BS $j$, and the SINR $\eta_{jk}^{\rm{ul}}$ is given by
\begin{eqnarray} \label{sinr_ul}
\begin{split}
\eta_{jk}^{\rm{ul}} &= {\frac{\tau_{jk}\left|{\bf g}_{jk}^H {\hat{\bf h}}_{jjk} \right|^2 }{\Eset \left\{ \tau_{jk}\left|{\bf g}_{jk}^H {\tilde{\bf h}}_{jjk} \right|^2 + \sum\limits_{\left(l,m\right) \ne \left(j,k \right)} \tau_{lm}\left|{\bf g}_{jk}^H {{\bf h}}_{jlm} \right|^2 +\sigma^2 \left\| {\bf g}_{jk} \right\|^2 \bigg|{\hat {\bf h}}_{(j)}\right\}}} \\
& = {\frac{\tau_{jk}{\bf g}_{jk}^H {\hat{\bf h}}_{jjk} {\hat{\bf h}}_{jjk}^H{\bf g}_{jk} }{{\bf g}_{jk}^H \left( \tau_{jk}{\bf C}_{jjk} + \sum\limits_{\left(l,m\right) \ne \left(j,k \right)} \tau_{lm}\left( {\hat{\bf h}}_{jlm}{\hat{\bf h}}_{jlm}^H +{\bf C}_{jlm} \right) + \sigma^2\right){\bf g}_{jk}}},
\end{split}
\end{eqnarray}
where $\Eset\{\cdot | {\hat {\bf h}}_{(j)}\}$ denotes the conditional expectation given all the estimated channels at BS $j$. Due to that the imperfectly estimated channels are available, the SE in~(\ref{rate_ul}) is achieved by treating ${\bf g}_{jk}^H {\hat{\bf h}}_{jjk}$ as the true channel, and treating uncorrelated interference and channel uncertainty as worst-case Gaussian noise~\cite{Hoydis2013}. Thus, $R_{jk}^{\rm{ul}}$ is a lower bound on the uplink ergodic capacity.

The second line of Eqn.~(\ref{sinr_ul}) shows that the uplink SINR takes the form of a generalized Rayleigh quotient. Therefore, a new multi-cell MMSE (M-MMSE) detector can be derived to maximize this SINR for given channel estimates:
\begin{equation} \label{detector}
{\bf{g}}_{jk}^{\rm{M-MMSE}} = \left( {\sum\limits_{l \in \cal L} {\sum\limits_{k = 1}^K \tau_{lk}{{\hat {\bf h}}_{jlk}{\hat {\bf h}}_{jlk}^H}}  + \sum\limits_{l \in \cal L} {\sum\limits_{k = 1}^K \tau_{lk} {{{\bf{C}}_{jlk}}} }  + {\sigma ^2}{{\bf{I}}_M}} \right)^{-1}{\hat {\bf h}}_{jjk}.
\end{equation}
As the name suggests, this detector (with an appropriate scaling) also minimizes the mean square error (MSE) in estimating $x_{jk}$~\cite{tse2005fundamentals}:
\begin{equation}
\Eset \left\{ \left|{\hat x}_{jk} -x_{jk} \right|^2 \big|{\hat {\bf h}}_{\left(j \right)}\right\}.
\end{equation}
By plugging~(\ref{errormatrix2}) and~(\ref{estimation2}) into~(\ref{detector}), the M-MMSE detector can also be expressed as
\begin{equation} \label{detector2}
{\bf{g}}_{jk}^{\rm{M-MMSE}}= {\left( {{{{\hat {\bf H}}}_{\mathcal{V},j}}{{\bf{\Lambda }}_j}{\hat {\bf H}}_{\mathcal{V},j}^H + \left( {{\sigma ^2} + {\varphi_j}} \right){{\bf{I}}_M}} \right)^{ - 1}}{\hat {\bf h}}_{jjk},
\end{equation}
where ${{\bf{\Lambda }}_j} = \sum\limits_{l \in L} {\sum\limits_{k = 1}^K \tau_{lk} p_{lk}d_j^2({\bf z}_{lk})} {{\bf{e}}_{{i_{lk}}}}{\bf{e}}_{{i_{lk}}}^H$ is a diagonal matrix, and its $i$th diagonal element ${\lambda _{ji}}$ depends on the large scale fading, the pilot and  payload power of the users that use the $i$th pilot sequence in $\cal V$. The scalar ${\varphi_j}$ is defined as ${\varphi_j} = \sum\limits_{l \in {\cal L}} {\sum\limits_{k = 1}^K {\tau_{lk}d_j({\bf z}_{lk}) (1-p_{lk}d_j({\bf z}_{lk}) \alpha_{ji_{lk}}B) }}$, where $\alpha_{ji_{lk}}$ is defined in Eqn.~(\ref{temp}).

To elaborate the advantages of our M-MMSE scheme, we compare it with some related work. First, the conventional single-cell MMSE (S-MMSE) detector from~\cite{Hoydis2013,Guo2014,Krishnan2014} is
\begin{equation} \label{smmse}
{\bf{g}}_{jk}^{\rm{S-MMSE}} = \left(  {\sum\limits_{m = 1}^K \tau_{jm}{{\hat {\bf h}}_{jjm}{\hat {\bf h}}_{jjm}^H} } + {\bf Z}_j + {\sigma ^2}{{\bf{I}}_M} \right)^{-1}{\hat {\bf h}}_{jjk},
\end{equation}
where inter-cell interference is either ignored by setting ${\bf Z}_j = \bf 0$ or only considered statistically as with
\begin{equation} \label{z}
{\bf Z}_j = \Eset\left\{ \sum\limits_{m=1}^K \tau_{jm}{{\tilde {\bf h}}_{jjm}{\tilde {\bf h}}_{jjm}^H} + \sum\limits_{l \ne j}\sum\limits_{m=1}^K \tau_{jm}{{ {\bf h}}_{jlm}{{\bf h}}_{jlm}^H}  \right\}.
\end{equation}

Notice that the S-MMSE detector in~(\ref{smmse}) is not a pure single-cell detector if ${\bf Z}_j$ in~(\ref{z}) is used, since statistical information about the multi-cell interfering channels is utilized in ${\bf Z}_j $. We refer to it as ``single-cell'' detector because it only utilizes the $K$ estimated channel directions from within the serving cell, and treats directions from other cells as uncorrelated noise. In comparison, all the $B$ available estimated directions in ${\hat {\bf H}}_{{\cal V},j}$ are utilized in our M-MMSE detector so that BS $j$ can actively suppress also parts of inter-cell interference when $B > K$. Therefore, our detector can actually maximize the SINR in~(\ref{sinr_ul}), while the S-MMSE can only do this in single-cell scenarios. The M-MMSE scheme can be seen as a coordinated beamforming scheme, but since there is no signalling between the BSs (BS $j$ estimates ${\hat {\bf H}}_{{\cal V},j}$ from the uplink pilots), the M-MMSE scheme is fully scalable.

Compared with the multi-cell MMSE scheme proposed in~\cite{Hien2012} and~\cite{Guo2013}, our detector is more practical and general. To begin with, power control and any fractional pilot reuse policy are supported in our scheme, which allows for an analysis based on a more flexible and practical network deployment. It is shown in~\cite{bjornson2014massive} that in massive MIMO systems, fractional pilot reuse is an important way to suppress pilot contamination and achieve high system SE. Furthermore, the uplink detector in~\cite{Guo2013} is based on the unrealistic assumption that perfect CSI is known at each BS, while imperfect channel estimation is accounted for in our detector. Thus the performance gains provided by our detector are actually achievable in practical systems. This makes our new M-MMSE detector the state-of-the-art method for massive MIMO detection. In Section~\ref{asymptotic analysis}, an explicit large-scale approximation expression of the SINR in~(\ref{sinr_ul}) is provided, which allows for simple performance analysis and the design of resource allocation schemes without time-consuming Monte Carlo simulation.

\subsection{Downlink Multi-cell MMSE Precoder} \label{mmse_precoder}
During the downlink payload data transmission, the received signal at user $k$ in cell $j$ is
\begin{equation}
y_{jk} = \sum\limits_{l \in {\cal L}} {\bf h}_{ljk}^H \sum\limits_{m=1}^K \sqrt{\varrho}_{lm}{\bf w}_{lm} s_{lm} + n_{jk},
\end{equation}
where ${\bf w}_{lm}\in \mathbb{C}^{M \times 1}$ is the precoder used by BS $l$ for user $m$ in its cell, $s_{lm} \sim {\cal{CN}} (0,1 )$ is the payload data symbol for user $m$ in cell $l$, $\varrho_{lm}$ is the corresponding downlink transmit power coefficient, and $n_{jk}\sim {\cal{CN}}(0,1 )$ is AWGN.

Recently, an uplink-downlink duality for massive MIMO systems was established in~\cite{bjornson2014massive} which proves that for a proper downlink power control, the uplink SEs can be achieved also in the downlink if each downlink precoder is a scaled version of the corresponding uplink detector. Since the M-MMSE detector proposed in the Subsection~\ref{mmse detector} is the state-of-the-art uplink method, we apply the same methodology for downlink precoding. The downlink M-MMSE precoder is constructed as
\begin{equation} \label{m_precoder}
{\bf w}_{jk}^{\rm{M-MMSE}} = \frac{ {\bf g}_{jk}^{\rm{M-MMSE}}}{\sqrt{\gamma_{jk}}},
\end{equation}
where $\gamma_{jk} = \Eset\{\|{\bf g}_{jk}^{\rm{M-MMSE}} \|^2 \}$ normalizes the average transmit power for the user $k$ in cell $j$ to $\Eset\{\| \sqrt{\varrho_{lm}}{\bf w}_{jk}^{\rm{M-MMSE}}s_{lm}\|^2 \}=\varrho_{lm}$. Since there are no downlink pilots in the TDD protocol, the users do not know the current channel but can learn their statistical equivalent channels, $ \sqrt{\varrho_{jk}} \Eset_{\{\bf h\}}\{{\bf h}_{jjk}^H {\bf w}_{jk}\}$, and the total interference variance. Consequently, a downlink SE
\begin{equation} \label{rate_dl}
R_{jk}^{\rm{dl}} = \zeta^{\rm{dl}} \left(1-\frac{B}{S} \right) \log_2\left( 1+ \eta_{jk}^{\rm{dl}}\right)
\end{equation}
can be achieved for user $k$ in cell $l$~\cite{Hoydis2013,bjornson2014massive}, where $\eta_{jk}^{\rm{dl}}$ is
\begin{equation}\label{sinr_dl}
\eta_{jk}^{\rm{dl}} = \frac{{\varrho_{jk}} \left| \Eset_{\left\{\bf h\right\}}\left\{{\bf h}_{jjk}^H {\bf w}_{jk} \right\}\right|^2 }{\sum\limits_{l \in {\cal L}}  \sum\limits_{m=1}^K {\varrho_{lm}} \Eset_{\left\{\bf h\right\}}\left\{ \left| {\bf h}_{ljk}^H {\bf w}_{lm}\right|^2 \right\} - {\varrho_{jk}} \left| \Eset_{\left\{\bf h\right\}}\left\{{\bf h}_{jjk}^H {\bf w}_{jk} \right\}\right|^2  + \sigma^2 }.
\end{equation}
This downlink SINR holds for any linear precoding scheme, and we omit the superscript ``M-MMSE'' of ${\bf w}_{jk}$ for brevity. The SE in~(\ref{rate_dl}) is achieved by treating $\Eset_{\{\bf h\}}\{{\bf h}_{jjk}^H {\bf w}_{jk} \}$ as the true channel, and treating interference and channel variations as worst-case uncorrelated Gaussian noise. Thus, $R_{jk}^{\rm{dl}}$ is a lower bound on the downlink ergodic capacity.

By utilizing all the available estimated directions, the M-MMSE precoder can suppress intra-cell interference and also reduce the interference caused to other cells, and thus a higher SINR can be expected by our precoder than conventional single-cell precoders, at least for an appropriate power control~\cite{bjornson2014massive}. In the next section, a large-scale approxmiation of the downlink SINR in~(\ref{sinr_dl}) is derived. In~\cite{Jose2011}, the authors also proposed a multi-cell MMSE precoder which brings a notable gain over single-cell processing, but it does not accounted for arbitrary or optimized pilot allocation. Moreover, no closed-form performance expression is provided in~\cite{Jose2011}.

Looking jointly at the uplink and downlink, the ergodic achievable SE for user $k$ in cell $j$ is
\begin{equation} \label{jointSE}
R_{jk} = \left(1-\frac{B}{S}\right)\left(\zeta^{\rm{ul}} \Eset_{\left\{{\hat{\bf h}}_{\left(j\right)}\right\}} \left\{\log_2\left(1+\eta_{jk}^{\rm{ul}} \right) \right\}+ \zeta^{\rm{dl}}\log_2\left(1+\eta_{jk}^{\rm{dl}} \right) \right) .
\end{equation}

\section{Asymptotic Analysis} \label{asymptotic analysis}
In this section, performance analysis is conducted for the proposed multi-cell MMSE scheme. Since the uplink SINR in~(\ref{sinr_ul}) depends on the stochastic channel estimates in each block, the uplink SE in~(\ref{rate_ul}) cannot be computed in closed form. Therefore, a deterministic equivalent expression for the SINR is computed instead which is tight in the large-system limit. A large-scale approximation of the downlink SINR is also provided. The large-system limit is considered, where $M$ and $K$ go to infinity while keeping ${K \mathord{\left/  {\vphantom {K M}} \right. \kern-\nulldelimiterspace} M} $ finite. In what follows, the notation $M \to \infty$ refers to $K$, $M \to \infty$ such that $\lim {\sup _M}{K \mathord{\left/  {\vphantom {K M}} \right. \kern-\nulldelimiterspace} M} < \infty $ and $\lim {\inf _M}{K \mathord{\left/  {\vphantom {K M}} \right. \kern-\nulldelimiterspace} M} >0$.\footnotemark{ }Since $B$ scales with $K$ for a fixed $\beta$, $\lim {\sup _M}{B \mathord{\left/  {\vphantom {K M}} \right. \kern-\nulldelimiterspace} M} < \infty $ and $\lim {\inf _M}{B \mathord{\left/  {\vphantom {K M}} \right. \kern-\nulldelimiterspace} M} >0$ also hold for $B$. The results should be understood in the way that, for each set of system dimension parameters $M$, $K$ and $B$, we provide large-scale approximative expressions for the uplink SINR and downlink SINR, and the expressions are tight as $M$, $K$ and $B$ grow large. The main feature is that they only depend on the large-scale fading, power control and pilot allocation, and can be computed efficiently without the need for time-consuming Monte Carlo simulations. In what follows, the notation $\xrightarrow[M \to \infty]{a.s.}$ denotes almost sure convergence of a stochastic sequence, and $\xrightarrow[M \to \infty]{}$ denotes convergence of a deterministic sequence.
\footnotetext{The limit superior of a sequence $x_n$ is defined by $\lim {\sup _n}{x_n}\triangleq \mathop {\lim }\limits_{n \to \infty } \left( {\sup \left\{ {{x_m}:m \geqslant n} \right\}} \right)$; the limit inferior is defined as $\lim{\inf_n}{x_n}\triangleq \mathop {\lim }\limits_{n \to \infty } \left( {\inf \left\{ {{x_m}:m \geqslant n} \right\}} \right)$.}

Before we continue with our performance analysis, two useful results from large random matrix theory are first recalled in the following subsection. All vectors and matrices should be understood as sequences of vectors and matrices of growing dimensions.

\subsection{Useful theorems} \label{theorems}
\begin{Theorem} (Theorem 1 in \cite{Wagner2012}): \label{theorem1}
Let ${\bf D} \in \Cset ^{M \times M}$ be deterministic and ${\bf H} \in \Cset ^{M \times B}$ be random with independent column vectors ${\bf h}_b \sim {\cal {CN}} (0, \frac{1}{M}{\bf R}_b )$. Assume that $\bf D$ and the matrices ${\bf R}_b( b=1,...,B)$, have uniformly bounded spectral norms (with respect to $M$). Then, for any $\rho > 0$,
\begin{equation}
\frac{1}{M} {\rm{tr}}\left({\bf D} \left( {\bf {HH}}^H +\rho {\bf I}_M \right)^{-1}\right) - \frac{1}{M} {\rm{tr}}\left({\bf D}{\bf T}\left( \rho \right)\right) \xrightarrow[M \to \infty]{a.s.} 0,
\end{equation}
where ${\bf T}( \rho ) \in \Cset^{M \times M}$ is defined as
\begin{equation}
{\bf T}\left( \rho \right) = \left( \frac{1}{M}\sum\limits_{b=1}^B \frac{{\bf R}_b}{1+\delta_b \left( \rho\right)} +\rho{\bf I}_M\right)^{-1}
\end{equation}
and the elements of ${\bm {\delta }}(\rho) \buildrel \Delta \over = {[ {{\delta _1}(\rho ),...,{\delta _B}(\rho)}]^T}$ are defined as $\delta_b(\rho)=\lim_{t \to \infty}\delta_b^{(t)}( \rho), b=1,...,B$, where
\begin{equation}
\delta_b^{\left(t \right)}\left( \rho \right) = \frac{1}{M} {\rm{tr}} \left( {\bf R}_b \left( \frac{1}{M} \sum\limits_{j=1}^B \frac{{\bf R}_j}{1+ \delta_j^{\left(t-1\right)} \left(\rho \right)} +\rho{\bf I}_N\right)^{-1}\right)
\end{equation}
for $t=1,2,\ldots,$ with initial values $\delta_b^{(0 )} = 1/\rho$ for all $b$.
\end{Theorem}
\begin{Theorem}(From \cite{Wagner2012}) \label{theorem2}
Let ${\bf {\Theta}} \in \Cset^{M \times M}$ be Hermitian nonnegative definite with uniformly bounded spectral norm (with respect to $M$). Under the same conditions on ${\bf D}$ and ${\bf H}$ as in Theorem~\ref{theorem1},
\begin{equation}
\frac{1}{M} {\rm{tr}}\left({\bf D} \left( {\bf {HH}}^H + \rho {\bf I}_M \right)^{-1} {\bf \Theta}\left( {\bf {HH}}^H + \rho {\bf I}_M \right)^{-1} \right)- \frac{1}{M} {\rm{tr}}\left({\bf D}{\bf T}'\left( \rho \right) \right) \xrightarrow[M \to \infty]{a.s.} 0
\end{equation}
where ${\bf T}'( \rho ) \in \Cset^{M \times M}$ is defined as
\begin{equation}
{\bf T}'\left( \rho \right) ={\bf T}\left( \rho \right) {\bf \Theta} {\bf T}\left( \rho \right) +{\bf T}\left( \rho \right) \frac{1}{M} \sum\limits_{b=1}^B \frac{{\bf R}_b \delta'_b\left(\rho \right)}{\left(1+ \delta_b\left(\rho \right)\right)^2}{\bf T}\left( \rho \right).
\end{equation}
${\bf T}( \rho)$ and ${\bm \delta}( \rho)$ are defined in Theorem~\ref{theorem1}, and ${\bm{\delta }}'( \rho)={\delta}'_1( \rho ),...,{\delta}'_Bt( \rho ) ]^T$ is calculated as
\begin{equation}
{\bm \delta}' \left( \rho  \right)= \left({\bf I}_B - {\bf J}\left( \rho \right) \right)^{-1} {\bf v}\left( \rho \right)
\end{equation}
where ${\bf J}( \rho)$ and $ {\bf v}( \rho )$ are defined as
\begin{equation}
\left[ {\bf J}\left( \rho \right)\right]_{bl} = {\frac{ \frac{1}{M} {\rm{tr}} \left({\bf R}_b {\bf T}\left( \rho \right) {\bf R}_l {\bf T}\left( \rho \right)\right)} {M \left(1+\delta_l\left( \rho \right) \right)^2 }},  1 \le b,l \le B, \\
\end{equation}
\begin{equation}
\left[ {\bf v}\left( \rho \right)\right]_{b} = \frac{1}{M}{\rm{tr}}\left({\bf R}_b{\bf T}\left( \rho \right) {\bf \Theta}{\bf T}\left( \rho \right)\right),  1 \le b \le B.
\end{equation}
\end{Theorem}
\vspace{-1ex}
\subsection{Large-scale Approximations of the SINRs with the M-MMSE scheme} \label{deterministic sinr}
In what follows, we derive the deterministic equivalent ${\bar \eta}_{jk}^{\rm{ul}}$ of ${\eta}_{jk}^{\rm{ul}}$ with the M-MMSE detector, and the large-scale approximation ${\bar \eta}_{jk}^{\rm{dl}}$ of ${\eta}_{jk}^{\rm{dl}}$ with the M-MMSE precoder, such that
\begin{equation}
{\bar \eta}_{jk}^{\rm{ul}}-{\eta}_{jk}^{\rm{ul}} \xrightarrow[M \to \infty]{a.s.} 0, \,\,\,\,\,\, {\bar \eta}_{jk}^{\rm{dl}}-{\eta}_{jk}^{\rm{dl}} \xrightarrow[M \to \infty]{}0.
\end{equation}
\begin{Theorem} \label{theorem3}
For the uplink M-MMSE detector in~(\ref{detector2}), we have ${{\bar \eta }_{jk}}^{\rm{ul}}-\eta_{jk}^{\rm{ul}}  \xrightarrow[M \to \infty]{a.s.}0$, where $\bar{\eta}_{jk}^{\rm{ul}}$ is given by
\begin{equation} \label{sinr_ul_determ}
{{\bar \eta }_{jk}^{\rm{ul}}} = \frac{\tau_{jk} p_{jk} d_j^2\left({\bf z}_{jk} \right){\delta _{jk}^2}}{{\delta_{jk}^2\sum\limits_{\left( {l,m} \right) \ne \left( {j,k} \right),{i_{_{lm}}} = {i_{jk}}} {\tau_{lm} p_{lm} d_j^2\left({\bf z}_{lm} \right)}  + \sum\limits_{{i_{_{lm}}} \ne {i_{jk}}} { \tau_{lm} d_j\left({\bf z}_{lm} \right)\frac{\mu_{jlmk}}{M}}  + \frac{{{\sigma ^2}}}{M}{\vartheta^{''}_{jk}}}},
\end{equation}
with
\begin{flalign}
& \delta_{jk}=\frac{1}{M} {\rm{tr}} \left({\tilde {\bf \Phi}}_{{\cal V},ji_{jk}} {{\bf T}}_j\right) & \nonumber
\end{flalign}
\begin{flalign}
&\mu_{jlmk}=\frac{1}{M}{\rm{tr}} \left({{{{\mathbf{ T}}}_{jk}^{'}}}\right) - p_{lm} d_j\left({\bf z}_{lm}\right)\lambda_{ji_{lm}}\vartheta_{jlmk}^{'}\vartheta_{jlm} \frac{2+\lambda_{ji_{lm}} \vartheta_{jlm} }{ \left(1+ \lambda_{ji_{lm}} \vartheta_{jlm}\right)^2 } & \nonumber
\end{flalign}
\begin{flalign}
&\vartheta_{jlm}=\frac{1}{M}{\rm{tr}}\left( {\tilde {\bf \Phi}}_{{\cal V},ji_{lm}}{\bf T}_j\right)& \nonumber
\end{flalign}
\begin{flalign}
&\vartheta_{jlmk}^{'}=\frac{1}{M}{\rm{tr}}\left( {\tilde {\bf \Phi}}_{{\cal V},ji_{lm}}{\bf T}_{jk}^{'}\right)& \nonumber
\end{flalign}
\begin{flalign}
&\vartheta_{jk}^{''}=\frac{1}{M}{\rm{tr}}\left( {\tilde {\bf \Phi}}_{{\cal V},ji_{jk}}{\bf T}_{jk}^{''}\right) &\nonumber
\end{flalign}
where
\begin{enumerate}
\item ${\bf T}_j = {\bf T}_j(\alpha)$ and ${\bm{\delta }}( \alpha ) \buildrel \Delta \over = {[ {{\delta _1},...,{\delta _B}}]^T}$ are given by Theorem~\ref{theorem1} for $\alpha = \frac{\sigma^2 + \varphi_j}{M}$ and ${\bf R}_b =\lambda_{jb}{ \tilde {\bf \Phi}}_{{\cal{V}},jb}$.  %${\bf D}={\bf \Phi}_{jjk}$, ${\bf S}={\bf 0}$,
\item ${\bf T}_{jk}^{'} = {\bf T}_{jk}^{'}(\alpha)$ and ${\bm{\delta }}'( \alpha)= [{\delta}'_1,...,{\delta}'_B ]^T$ are given by Theorem~\ref{theorem2} for $\alpha = \frac{\sigma^2 +\varphi_j}{M}$, ${\bf \Theta}={\tilde{\bf \Phi}}_{{\cal V},ji_{jk}}$, and ${\bf R}_b =\lambda_{jb}{\tilde {\bf \Phi}}_{{\cal{V}},jb}$. %${\bf D}={\bf \Phi}_{jlm} $, ${\bf S}={\bf 0}$,
\item ${\bf T}_{jk}^{''}={\bf T}_{jk}^{''}(\alpha)$ and ${\bm{\delta }}'(\alpha )= [{\delta}'_1,...,{\delta}'_B]^T$ are given by Theorem~\ref{theorem2} for $\alpha = \frac{\sigma^2 + \varphi_j}{M}$, ${\bf \Theta}={\bf I}_M $, and ${\bf R}_b =\lambda_{jb}{\tilde {\bf \Phi}}_{{\cal{V}},jb}$.  %${\bf D}={\bf \Phi}_{jjk}$, ${\bf S}={\bf 0}$,
\end{enumerate}
\end{Theorem}
\noindent\emph{Proof:} See Appendix~\ref{sec:proof_thr3}. \hfill{$\blacksquare$}
\begin{Theorem} \label{theorem4}
For the downlink M-MMSE precoder in~(\ref{m_precoder}), we have ${{\bar \eta }_{jk}}^{\rm{dl}} - \eta_{jk}^{\rm{dl}}  \xrightarrow[M \to \infty]{}0$, where $\bar{\eta}_{jk}^{\rm{dl}}$ is given by
\begin{equation} \label{sinr_dl_determ}
{{\bar \eta }_{jk}}^{\rm{dl}}= \frac{\varrho_{jk} p_{jk} d_j^2\left({\bf z}_{jk} \right)\frac{\delta _{jk}^2}{\vartheta_{jk}^{''}}}{{p_{jk}\sum\limits_{\left( {l,m} \right) \ne \left( {j,k} \right),{i_{_{lm}}} = {i_{jk}}} {\varrho_{lm}d_l^2\left({\bf z}_{jk} \right)\frac{\delta_{lm}^2}{\vartheta_{lm}^{''}}}  + \sum\limits_{{i_{_{lm}}} \ne {i_{jk}}} { \varrho_{lm} d_l\left({\bf z}_{jk} \right)\frac{\mu_{ljkm}}{M \vartheta_{lm}^{''}}}  + \frac{{{\sigma ^2}}}{M}}},
\end{equation}
where $\delta_{lm}$, $\mu_{ljkm}$ and $\vartheta_{lm}^{''}$ are given in Theorem~\ref{theorem3}.
\end{Theorem}
\noindent\emph{Proof:} See Appendix~\ref{sec:proof_thr4}. \hfill{$\blacksquare$}

By utilizing Theorem~\ref{theorem3} and~\ref{theorem4}, the ergodic SEs $R_{jk}^{\rm{ul}}$ in~(\ref{rate_ul}) and $R_{jk}^{\rm{dl}}$ in~(\ref{rate_dl}), after dropping the prelog factor $(1-\frac{B}{S})$, converge to ${\bar R}_{jk}^{\rm{ul}}=\log_2 (1+{\bar \eta}_{jk}^{\rm{ul}})$ and  ${\bar R}_{jk}^{\rm{dl}}=\log_2 (1+{\bar \eta}_{jk}^{\rm{dl}})$ in the large-system limit, respectively. Therefore, a large-scale approximation of the joint ergodic SE in~(\ref{jointSE}) is provided by $(1-\frac{B}{S})(\zeta^{\rm{ul}} {\bar R}_{jk}^{\rm{ul}}+\zeta^{\rm{dl}}{\bar R}_{jk}^{\rm{dl}})$. This approximation is easy to compute and only depends on the large-scale fading, power control and pilot allocation. As shown in Section~\ref{simulation}, this large-scale approximation is very accurate also at small system dimensions.
\vspace{-1ex}
\subsection{The Uplink and Downlink Duality for the M-MMSE scheme} \label{up_down_duality}
It is pointed out in~\cite{bjornson2014massive} that when the precoder is a scaled version of the detector, like~(\ref{m_precoder}) in our case, the same per user SEs as in the uplink can be achieved in the downlink by properly selecting the downlink payload power. We establish this uplink-downlink duality for our M-MMSE scheme, using the large-scale SINR approximations given by Theorem~\ref{theorem3} and Theorem~\ref{theorem4}.
\begin{Theorem} \label{theorem5}
For the proposed M-MMSE scheme, if ${\bar \eta}_{jk}^{\rm{ul}}$ in~(\ref{sinr_ul_determ}) is achievable in the uplink for user $k$ in cell $j$, then a downlink power control policy $\{\varrho_{jk}\}$ can be obtained by transforming the corresponding uplink power $\{\tau_{jk}\}$ according to Eqn.~(78), such that $\sum\limits_{j\in {\cal L}} \sum\limits_{k=1}^K \tau_{jk} = \sum\limits_{j\in {\cal L}} \sum\limits_{k=1}^K \varrho_{jk}$ and that the same SE is achieved in the downlink, i.e., ${\bar \eta}_{jk}^{\rm{dl}}={\bar \eta}_{jk}^{\rm{ul}}$.
\end{Theorem}
\noindent\emph{Proof:} See Appendix~\ref{sec:proof_thr5}. \hfill{$\blacksquare$}

Note that Theorem~\ref{theorem5} establishes the duality for the large-scale SINR approximations, instead of the real SINRs. However, since the approximations are very accurate even for small system dimensions, Theorem~\ref{theorem5} provides a powerful tool to obtain a judicious downlink power allocation whenever the same SEs are desired in both the uplink and downlink.

\section{Iterative Power Control}\label{sec:powercontrol}
The large-scale approximations of the uplink and downlink SINRs given in Theorem~\ref{theorem3} and Theorem~\ref{theorem4} not only enable us to evaluate the system performance without time-consuming Monte Carlo simulation, but they also enable us to improve the system performance by optimizing key system parameters based on only large-scale fading. In this section, we consider optimizing the uplink payload transmit power jointly for the multi-cell network to maximize the weighted uplink sum SE. Since the downlink payload power can be obtained according to Theorem~\ref{theorem5}, the optimized uplink SEs can be achieved also in the downlink using the same total transmit power. The effectiveness of our proposed power control algorithm is testified in Section~\ref{simulation}.

\subsection{Joint Uplink Power Control for Weighted Uplink Sum SE Maximization}
The power control for sum SE maximization has been widely studied in cellular networks~\cite{chiang2008power,Gesbert2007adaptation,Luo2006an,Chiang2005balancing,Paschalidis2007,kumaran2006uplink,charafeddine2007rate,bjornson2013Optimal}, and here we consider this sum SE metric for the proposed M-MMSE detector. Using the same notations of ${\bf D}$, ${\bf F}$ and $\bm {\tau}$ defined in Appendix~\ref{sec:proof_thr5}, and define the vector ${\bm r} =\left[{\bar \eta}_{11}^{\rm{ul}},\ldots, {\bar \eta}_{LK}^{\rm{ul}} \right]^T \in \Rset^{LK \times 1}$, then the uplink SINR approximation in~(\ref{sinr_ul_determ}) can be expressed as
\begin{equation} \label{r1}
r_l = {\bar \eta}_{jk}^{\rm{ul}} = \frac{\tau_l D_{l,l}}{\left({\bf F}{\bm \tau} \right)_l + \frac{\sigma^2}{M}},
\end{equation}
where $(\cdot)_l$ denotes the $l$th element of the corresponding vector and $l = k+(j-1)K$. Using the notation in~(\ref{r1}), we want to find the power control that maximizes the weighted SE as
\begin{equation}
\begin{aligned}
& {\mathcal{P}}: \quad \underset{\bm \tau} {\text{maximize}} \quad \sum\limits_{l=1}^{LK} \xi_{l}\log_2\left( 1+ r_l\right) \nonumber \\
%\begin{array}{r@{\quad}r@{}l@{\quad}l}
& s.t. \quad 0 \le \tau_l  \le P_{max},\,\,\, \forall l,
\end{aligned}
\end{equation}
where $P_{max}$ is the maximum radiated transmit power of each user and $\xi_l > 0$ is the weight for the corresponding user. All $\xi_l = 1$ corresponds to conventional sum SE maximization, while other values can be used to enforce some fairness. However, as proved in~\cite{Luo2008}, power control problems for sum SE maximization are strongly NP-hard. Thus lower bounding of $\log_2( 1+ r_l)$ by $\log_2(r_l)$ is often used to approximate ${\cal P}$ as ${\cal P}_1$~\cite{Chiang2007,Tan2009}:
\begin{figure}[h]
\vspace{-2ex}
\begin{minipage}[!t]{0.4\linewidth}
%\begin{align*}
\begin{equation}
\begin{aligned}
&{\mathcal{P}}_1: \quad \underset{\bm \tau} {\text{maximize}} \quad \sum\limits_{l=1}^{LK} \xi_{l}\log_2\left( r_l\right) \nonumber  \\
& s.t. \quad 0 \le \tau_l \le P_{max}\,\,\, \forall l.
\end{aligned}
\end{equation}
%\end{array}
%\end{align*}
\end{minipage}
\hfil
\begin{minipage}[!t]{0.52\linewidth}
%\begin{align*}
\begin{equation}
\begin{aligned}
{\mathcal{P}}_2: &\quad \underset{{\bm \tau},{\bf q}} {\text{maximize}} \quad \prod\limits_{l=1}^{LK} q_l \nonumber\\
s.t. &\quad q_l^{\frac{1}{\xi_l}}\left( \sum\limits_{j=1}^K {F_{lj}\tau_j + \frac{\sigma^2}{M}} \right)\tau_l^{-1} D_{l,l}^{-1} \le 1,\,\,\, \forall l, \nonumber\\
     &\quad 0 \le \tau_l\le P_{max},\,\,\, \forall l.
\end{aligned}
\end{equation}
%\end{align*}
\end{minipage}
\vspace{-4ex}
\end{figure}

For fixed $\bf F$ and $\bf D$, by introducing the auxiliary vector $\bf q$ with its $l$th element $q_l \le r_l^{\xi_l}$, problem ${\cal P}_1$ can be turned into the geometric programming (GP) problem ${\cal P}_2$ above. The optimal solution of ${\cal P}_2$ can be obtained numerically, for example, using the convex optimization toolbox in MATLAB. A low-complexity fixed point iteration method is also proposed in~\cite{Tan2009} to solve problems of the same type as ${\cal P}_2$. With our notation, the power coefficient ${\tau}_l$ is updated as
\begin{equation} \label{update}
{\tau}_l\left( t+1\right) =\min \left\{{\xi_l}\bigg/\left({\sum\limits_{j=1}^{LK}{\frac{ \xi_jF_{j,l}\left( t\right)r_j\left( t\right)}{D_j\left( t\right) \tau_j\left( t\right)}} }\right),P_{max}\right\},
\end{equation}
where $t$ is the iteration index in the fixed point algorithm, for $t = 0, 1, \ldots$. It is proved in~\cite{Tan2009} that starting from the initial point $\tau_l ( 0)= P_{max}$ for all $l$, the above algorithm converges at a geometric rate to the optimal solution of ${\cal P}_1$ (for fixed $\bf F$ and $\bf D$).

In our case, however, $\bf F$ and $\bf D$ are not fixed since $\delta_{jk}$, $\mu_{jlmk}$ and $\vartheta_{jk}^{''}$ will change as $\tau_l$ changes. Hence, ${\cal P}_2$ in our work is not a pure GP. Therefore, Algorithm~\ref{alg:alg1} is proposed to iterate between solving ${\cal P}_2$ for fixed $\bf F$ and $\bf D$, and updating $\bf F$ and $\bf D$ using the current $\bm \tau$.
%\vspace{-1ex}
\begin{algorithm}[h]
\caption{\bf{: Approximated Sum SE Maximization Power Control Algorithm}}
\label{alg:alg1}
\begin{algorithmic}[1]
\STATE Initialize: $\tau_l ( 0)= P_{max}$ for all $l$, $t=0$ and select $\epsilon>0$.
\STATE Calculate ${\bf F} ( t)$, $\bf D( t)$ and $R( t) = \sum\limits_{l=1}^{LK} \xi_{l}\log_2( r_l)$ using ${\bm \tau} ( t)$.
\STATE Update ${\bm \tau}(t+1)$ by~(\ref{update}), and calculate $R( t+1)$ based on the newly updated ${\bm \tau}(t+1)$ and the ${\bf F} ( t)$ and ${\bf D}(t)$ in step 2, until $\left|R( t+1)-R(t) \right| \le \epsilon$.
\STATE Update the time slot index $t$ with $t+1$.
\STATE Repeat step 2 -- 4 until $R( t)$ converges.
\end{algorithmic}
\end{algorithm}
%\vspace{-3ex}

In step 3, the matrices $\bf F$, $\bf D$, the current power $\tau_j$ and the SINR $r_j$ of all users in the network are needed at each BS. Thus Algorithm~\ref{alg:alg1} involves some information exchange among the BSs. However, since the asymptotic approximation only depends on long-term parameters, the information exchange overhead is much smaller than if the sum SE would be maximized in every coherence block based on the current small-scale fading. Moreover, the proposed algorithm only involves simple calculations and converges quickly, thus it is of low complexity. Since the convergence has been proved in~\cite{Tan2009} for fixed $\bf F$ and $\bf D$, and we improve them in each iteration, our algorithm converges to some local optimal solution of ${\cal P}_1$.

\section{Simulation Results} \label{simulation}
In this section, we illustrate the analytical contributions by simulation results for a symmetric hexagonal network topology. We apply the classic 19-cell-wrap-around structure to avoid edge effects and guarantee the consistent simulated performance for all cells; see Fig.~1. Each hexagonal cell in the network has a radius of $r = 500$ meters, and is surrounded by 6 interfering cells in the first tier and 12 in the second tier. To achieve a symmetric pilot allocation in this hexagonal cellular network, the pilot reuse factor can be $\beta \in \{1,3,4,7\}$ as shown in Fig. 1. For each pilot reuse policy, the same subset of pilots are allocated to the cells with the same color, and pilots in each cell are allocated randomly to the users.
\begin{figure}[H]
\begin{minipage}[t]{0.245\textwidth}
\centering
\scalebox{0.18}{\includegraphics{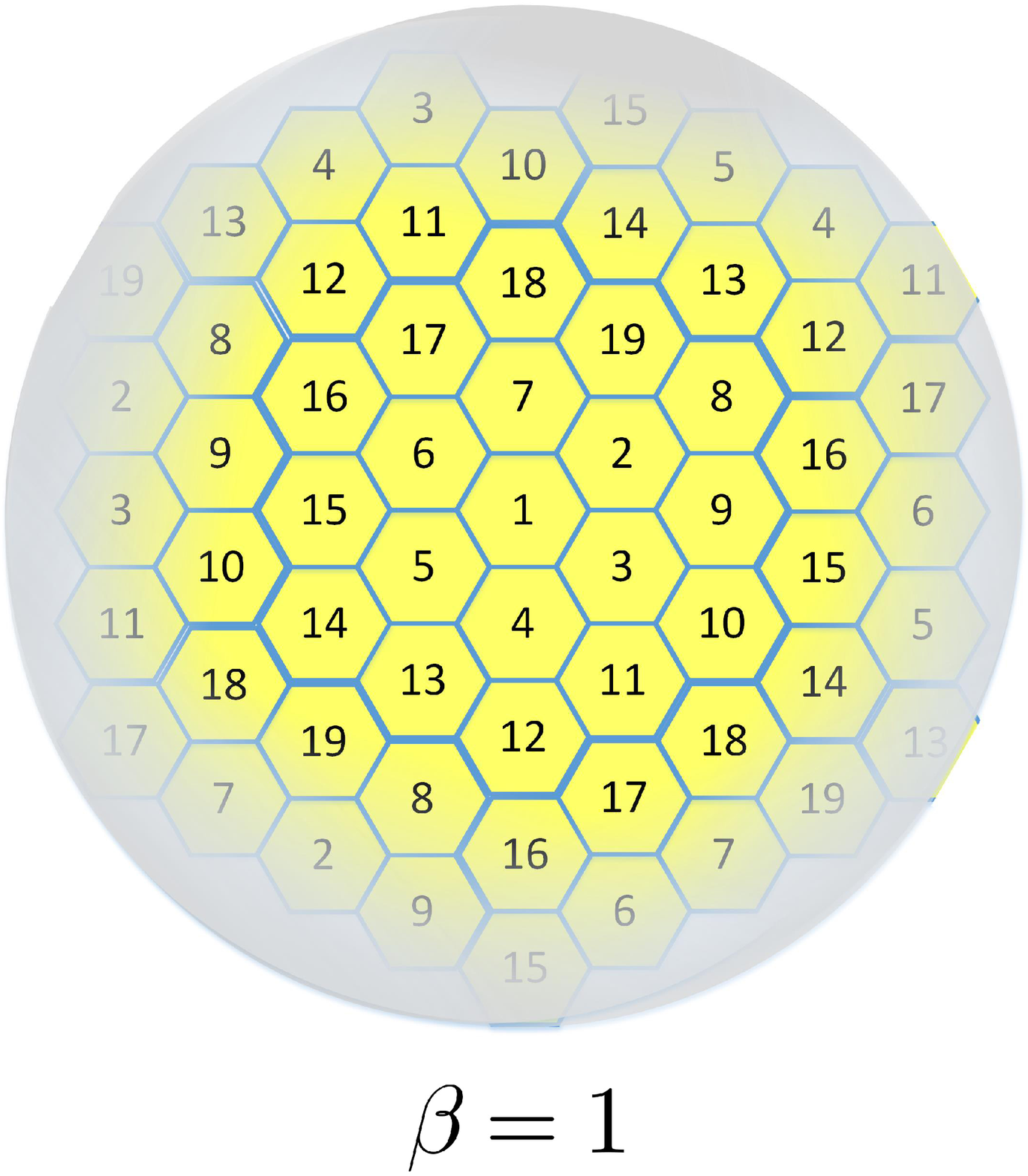}}  % use channel inversion power control for UL payload and p_max for DL payload
%\caption{$\beta = 1$.}
 \label{network}
\end{minipage}
\vspace{-1ex}
\begin{minipage}[t]{0.245\textwidth}
\centering
\scalebox{0.18}{\includegraphics{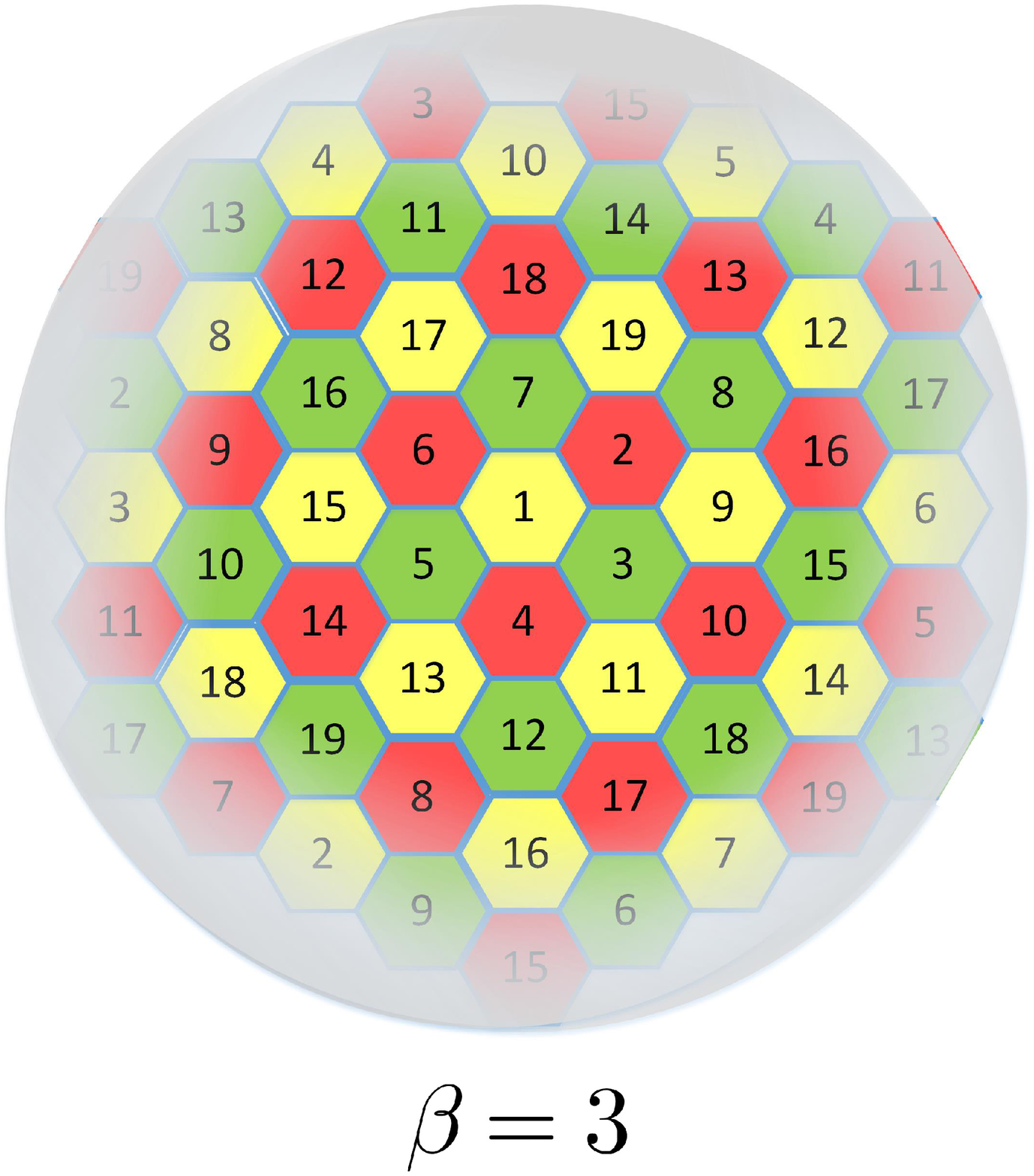}}
%\caption{$\beta = 3$.
\end{minipage}
\vspace{-1ex}
\begin{minipage}[t]{0.245\textwidth}
\centering
\scalebox{0.18}{\includegraphics{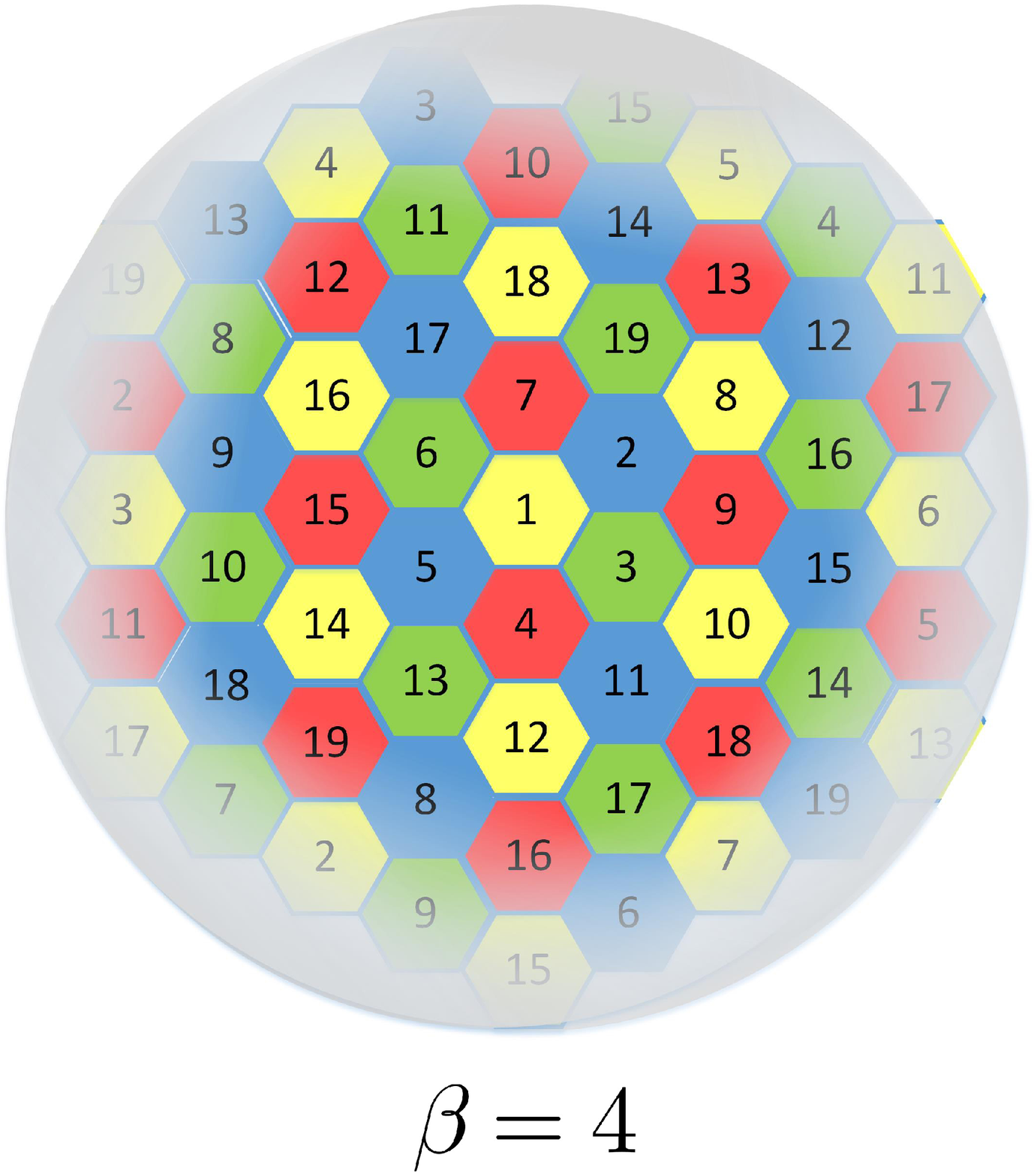}}
%\caption{$\beta = 4$.}
\end{minipage}
\vspace{-1ex}
\begin{minipage}[t]{0.245\textwidth}
\centering
\scalebox{0.18}{\includegraphics{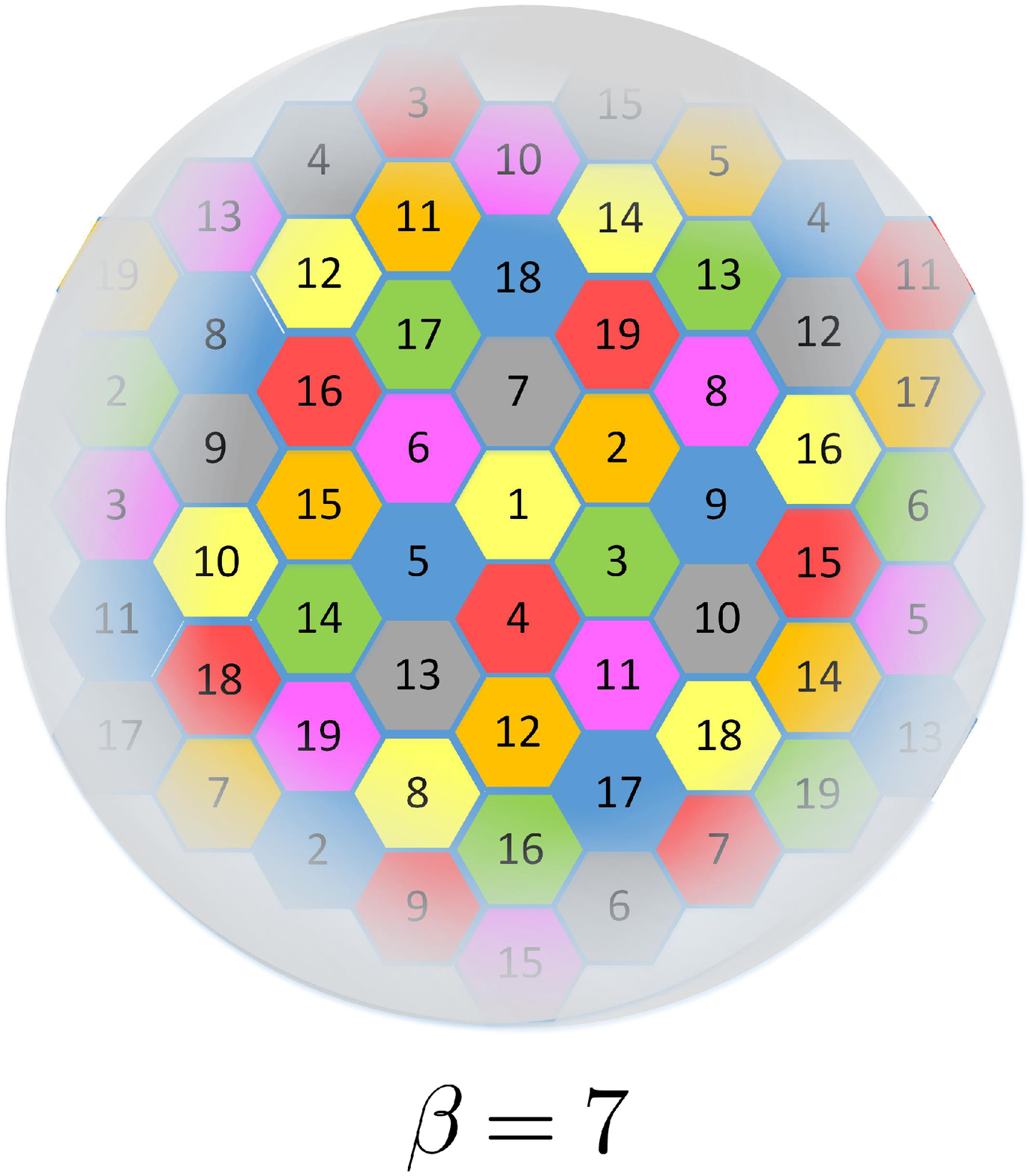}}
%\caption{$\beta = 7$.}
\end{minipage}
\vspace{-2ex}
\caption{The 19-cell-wrap-around hexagonal network topology for $\beta=1$, $\beta=3$, $\beta=4$ and $\beta=7$.}
\end{figure}
\vspace{-2ex}

The user locations are generated independently and uniformly at random in the in cells, but the distance between each user and its serving BS is at least $0.14r$. For each user location ${\bf z} \in \Rset^{2}$, a classic pathloss model is considered, where the variance of the channel attenuation is $d_j( {\bf z} ) = \frac{C_{(\bf z)}}{\| {\bf z} - {\bf b}_j\|^{\kappa}}$. The vector ${\bf b}_j \in \Rset^{2}$ is the location of the BS in cell $j$, $\kappa$ is the pathloss exponent, and $\|\cdot \|$ denotes the Euclidean norm. $C_{(\bf z)}>0$ is independent shadow fading for some user location $\bf z$ with $10\log_{10}(C_{(\bf z)})\sim {\cal N}(0,\sigma^2_{sf})$. In the simulation, we assume $\kappa = 3.7$, $\sigma_{sf}^2 = 5$ and the coherence block length $S= 1000$.\footnotemark{}
\footnotetext{This coherence block can, for example, have the dimensions of $T_c =10\,{\rm{ms}}$ and $W_c = 100\,{\rm{kHz}}$.}

\subsection{Benefits of the proposed M-MMSE scheme}\label{subA}
In this subsection, we show the benefits of our M-MMSE scheme over the conventional alternatives. Statistical channel inversion power control is applied to both pilot and uplink payload data, i.e., $p_{lk} =\tau_{lk}= \frac{\rho}{d_l({\bf z}_{lk})}$~\cite{bjornson2014massive}. Thus during the uplink phase, the average effective channel gain between users and their serving BSs is constant: $\Eset\{p_{lk}\|{\bf h}_{llk}\|^2\}=\Eset\{\tau_{lk}\|{\bf h}_{llk}\|^2\} = M\rho$. Then the average uplink SNR per antenna and user at its serving BS is $\rho/{\sigma^2}$. This is a simple but effective policy to avoid near-far blockage and, to some extent, guarantee a uniform user performance in the uplink. For downlink payload data transmission, the transmit power $\varrho_{lk}$ is selected according to Theorem~\ref{theorem5} to achieve the same downlink SE at each user as in the uplink. %the uplink-downlink duality discussion in Subsection~\ref{up_down_duality}.
In our simulation, $\rho/{\sigma^2}$ is set to 0 dB to allow for decent channel estimation accuracy, and the time proportions for the uplink and downlink are set to $\zeta^{\rm{ul}}=\zeta^{\rm{dl}}=\frac{1}{2}$.

\begin{figure}[tbp]
\begin{minipage}[t]{0.46\linewidth}
\centering
\scalebox{0.47}{\includegraphics{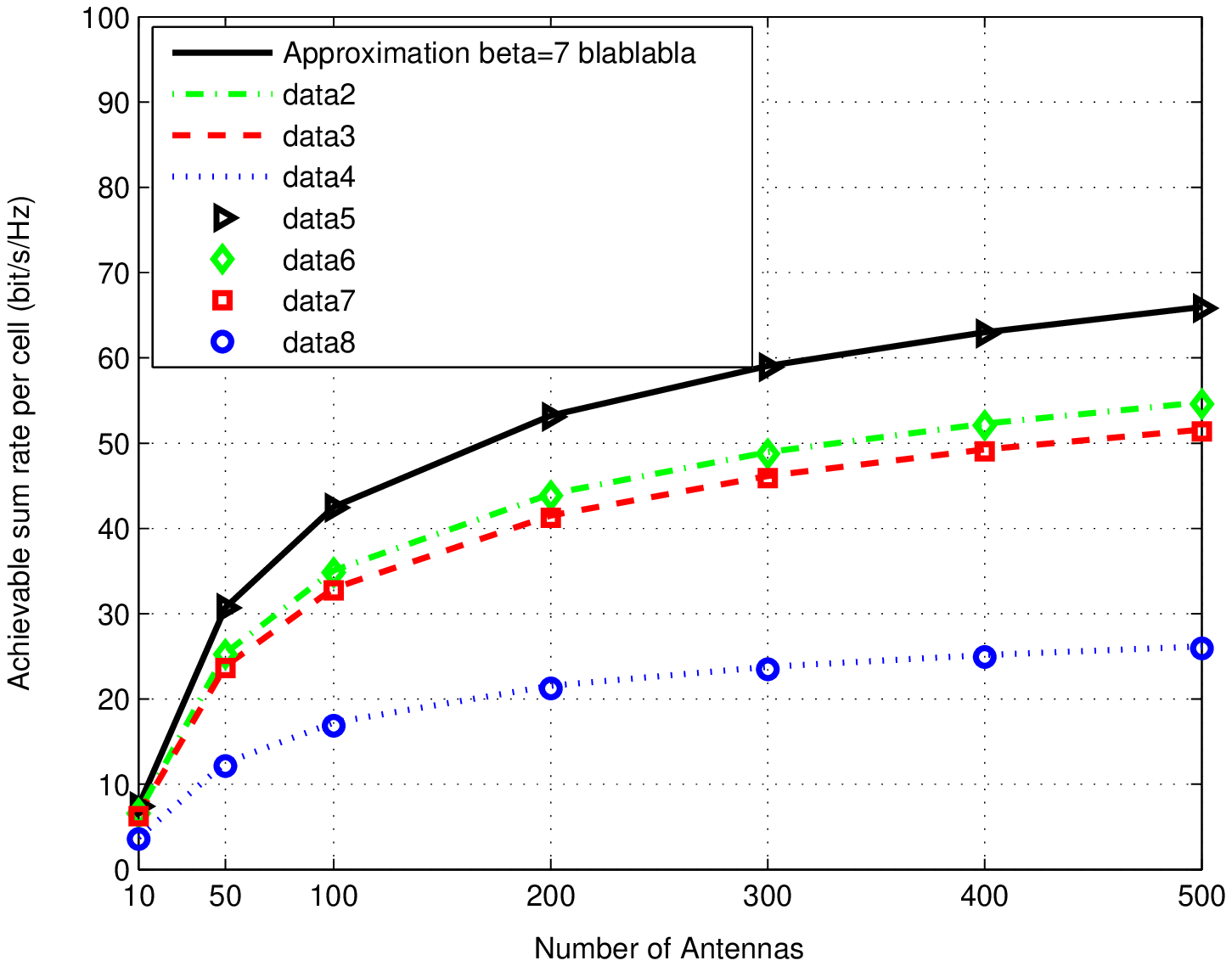}}
\caption{Achievable sum SE as a function of the number of antennas $M$, for $\beta\in\left\{1,3,4,7 \right\}$ and $K=10$.}
\label{inst_determ}
\end{minipage}
\hfil
\begin{minipage}[t]{0.46\linewidth}
\centering
\scalebox{0.47}{\includegraphics{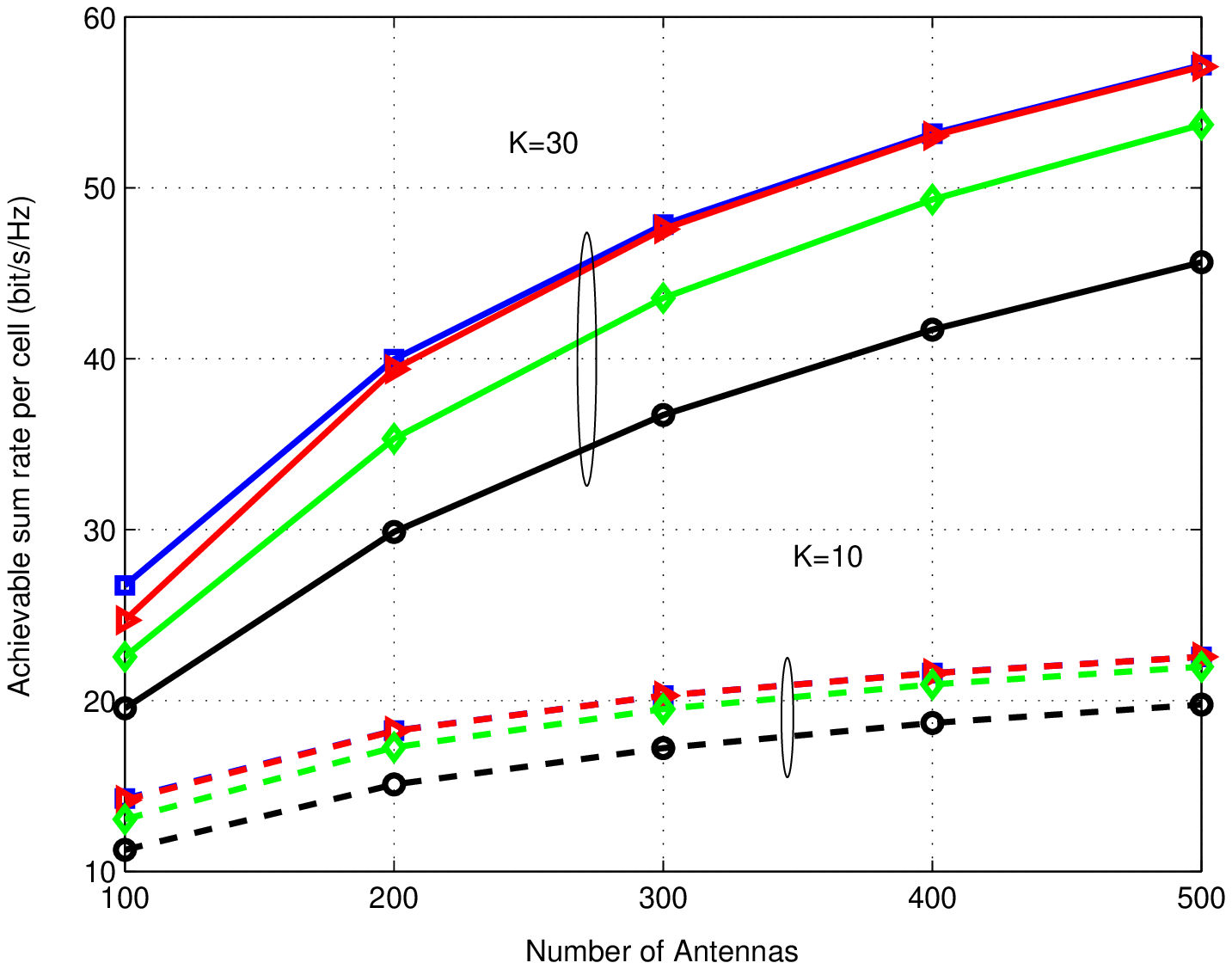}}
\caption{Achievable sum SE of M-MMSE (squares), M-ZF (triangles), S-MMSE (diamonds) and M-MF (circles) with $\beta =1$, $K=10$ and $K=30$.}
\label{sumrate_beta1}
\end{minipage}
\vspace{-2ex}
\end{figure}

To verify the accuracy of the large-scale approximations from Section~\ref{asymptotic analysis}, 10000 independent Monte-Carlo channel realizations are generated to numerically calculate the joint achievable SE in~(\ref{jointSE}). The numerical results and their large-scale approximations from Theorem~\ref{theorem3} and Theorem~\ref{theorem4} are shown in Fig.~\ref{inst_determ}. As seen from Fig.~\ref{inst_determ}, the achievable sum SE per cell increases monotonically with $\beta$ for the considered range of values. This is due to the following two properties. Firstly, a larger $\beta$ results in a lower level of pilot contamination, contributes to a higher channel estimation accuracy, and thereby increases the achievable SE. Secondly, a larger $\beta$ indicates more available estimated channel directions in the construction of the M-MMSE detector and precoder, thus a higher inter-cell interference suppression can be achieved. Moreover, Fig.~\ref{inst_determ} shows that the numerical results and the large-scale approximations match very well, even for small $M$ and small $K$.
% figure 1: rate of multicell mmse with different beta
\psfrag{Approximation beta=7 blablabla}{\Large {Approximation $\beta = 7$}}
\psfrag{data2}{\Large {Approximation $\beta = 4$}}
\psfrag{data3}{\Large {Approximation $\beta = 3$}}
\psfrag{data4}{\Large {Approximation $\beta = 1$}}
\psfrag{data5}{\Large {Simulation $\beta = 7$}}
\psfrag{data6}{\Large {Simulation $\beta = 4$}}
\psfrag{data7}{\Large {Simulation $\beta = 3$}}
\psfrag{data8}{\Large {Simulation $\beta = 1$}}
\psfrag{Number of Antennas}[][cb]{\Large {Number of Antennas}}
\psfrag{Achievable sum rate per cell (bit/s/Hz)}[][]{\Large{Achievable sum SE per cell (bit/s/Hz)}}

\psfrag{10}[][]{\Large {10}}
\psfrag{50}[][]{\Large {50}}
\psfrag{100}[][]{\Large {100}}
\psfrag{200}[][]{\Large {200}}
\psfrag{300}[][]{\Large {300}}
\psfrag{400}[][]{\Large {400}}
\psfrag{500}[][]{\Large {500}}

\psfrag{0}[][l]{\Large {0}}
\psfrag{20}[][l]{\Large {20}}
\psfrag{30}[][l]{\Large {30}}
\psfrag{40}[][l]{\Large {40}}
\psfrag{50}[][l]{\Large {50}}
\psfrag{60}[][l]{\Large {60}}
\psfrag{70}[][l]{\Large {70}}
\psfrag{80}[][l]{\Large {80}}
\psfrag{90}[][l]{\Large {90}}
\psfrag{100}[][l]{\Large {100}}
\psfrag{110}[][l]{\Large {110}}
\psfrag{120}[][l]{\Large {120}}
\psfrag{140}[][l]{\Large {140}}

% figure 1: rate comparison with beta =1
\psfrag{K=30}{\Large {$K=30$}}
\psfrag{K=10}{\Large {$K=10$}}

% figure 3: rate comparison with beta =4
\psfrag{K=30}{\Large {$K=30$}}
\psfrag{K=10}{\Large {$K=10$}}
\psfrag{Number of Antennas}[][cb]{\Large {Number of Antennas}}
\psfrag{Achievable Rate (bit/s/Hz)}[][]{\Large{Achievable Rate (bit/s/Hz)}}
\psfrag{500}[][]{\Large {500}}
\psfrag{200}[][]{\Large {200}}
\psfrag{300}[][]{\Large {300}}
\psfrag{400}[][]{\Large {400}}
\psfrag{100}[][]{\Large {100}}
\psfrag{110}[][]{\Large {110}}
\psfrag{120}[][]{\Large {120}}
\psfrag{140}[][]{\Large {140}}
\psfrag{160}[][]{\Large {160}}
\psfrag{180}[][]{\Large {180}}
% figure 4: beta = 7
\psfrag{data27}{\Large {Multicell MF}}
\psfrag{data37}{\Large {Multicell ZF}}
\psfrag{data47}{\Large {Singlecell MMSE}}
\psfrag{2}[][l]{\Large {2}}
\psfrag{3}[][l]{\Large {3}}
\psfrag{4}[][l]{\Large {4}}
\psfrag{5}[][l]{\Large {5}}
\psfrag{6}[][l]{\Large {6}}
\psfrag{7}[][l]{\Large {7}}
\begin{figure}[tbp]
\begin{minipage}[t]{0.46\linewidth}
\centering
\scalebox{0.47}{\includegraphics{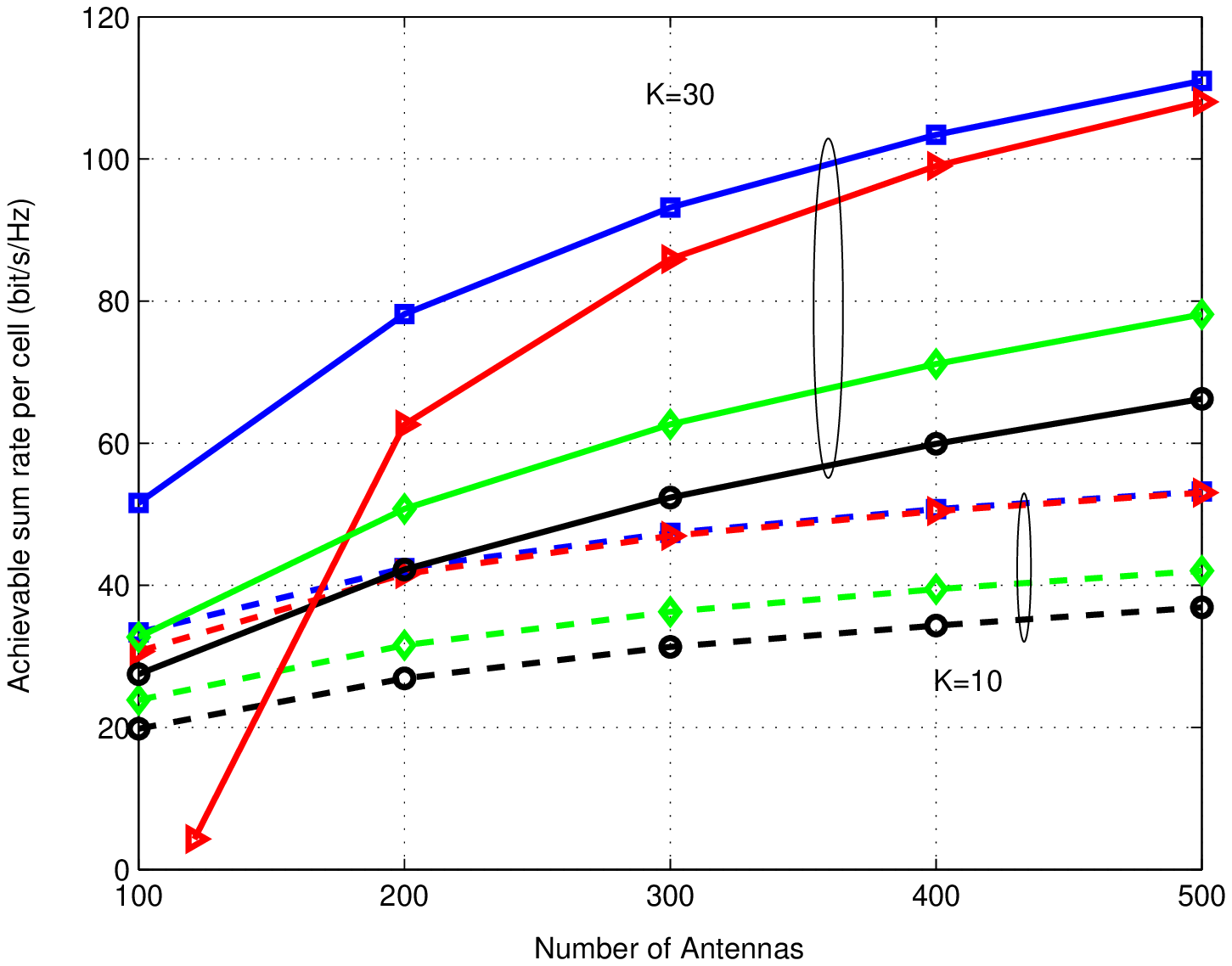}}
\caption{Achievable sum SE of M-MMSE (squares), M-ZF (triangles), S-MMSE (diamonds) and M-MF (circles) with $\beta =4$, $K=10$ and $K=30$.}
\label{sumrate_beta4}
\end{minipage}
\hfil
\begin{minipage}[t]{0.46\linewidth}
\centering
\scalebox{0.47}{\includegraphics{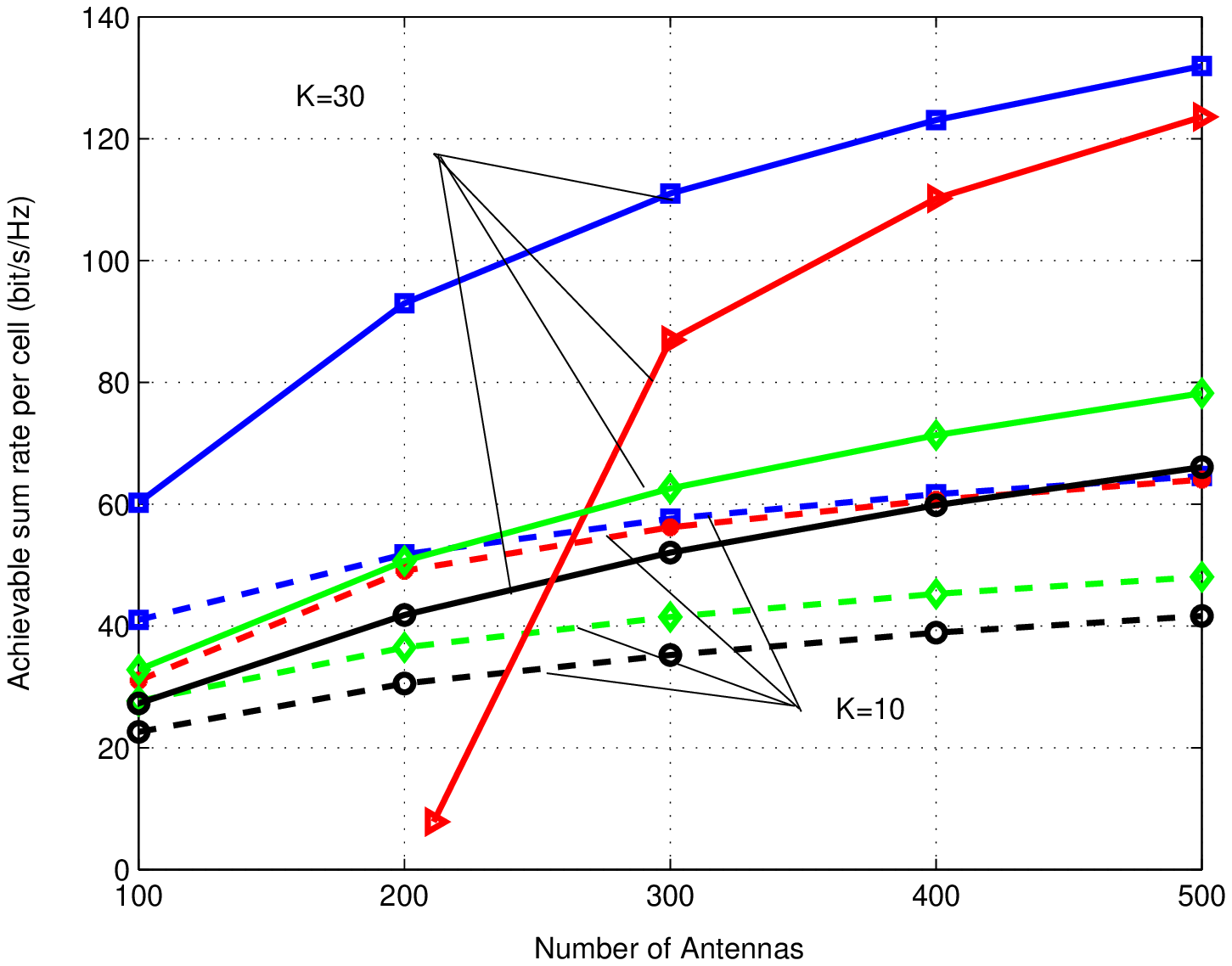}}
\caption{Achievable sum SE of M-MMSE (squares), M-ZF (triangles), S-MMSE (diamonds) and M-MF (circles) with $\beta =7$, $K=10$ and $K=30$.}
\label{sumrate_beta7}
\end{minipage}
\vspace{-3ex}
\end{figure}

To show explicitly the advantages of our M-MMSE scheme, simulation results for the matched filter (MF) from~\cite{Marzetta10}, the multi-cell ZF (M-ZF) scheme from~\cite{bjornson2014massive}, and the S-MMSE scheme from~(\ref{smmse}) are provided for comparison. The same downlink power acquisition from Theorem~\ref{theorem5} and normalization from~(\ref{m_precoder}) are applied for all precoders. Notice that $M-\beta K>0$ is needed for the M-ZF scheme, thus the minimum value of $M$ for the M-ZF is $\beta K + 1$. Simulation results are shown in Figs.~\ref{sumrate_beta1} --~\ref{sumrate_beta7} for $\beta=1$, $\beta=4$ and $\beta=7$, respectively. The MF scheme always achieves the lowest performance since it does not suppress any interference. Compared to the S-MMSE, our proposed M-MMSE always achieves a higher sum SE, and the advantage becomes more significant as $\beta$ and/or $K$ increases. For $\beta=4$ and $M = 200$, the SE of M-MMSE are 31\% and 53\% higher than those of S-MMSE for $K = 10$ and $K = 30$, respectively. For $\beta=7$, the gains increase to 42\% and 82\% for $K = 10$ and $K = 30$, respectively. The advantage of the M-MMSE over the M-ZF is only minor for small $\beta$ and small $K$, but the gain becomes notable as $\beta$ and $K$ grow. Since the complexity of our M-MMSE scheme is the same as for the M-ZF, and the M-ZF can sometimes achieve very low SE for small $M$, in general our M-MMSE scheme is the better choice if high system SE is desirable.

\subsection{Effectiveness of the joint power control scheme}
In this subsection, the effectiveness of the power control scheme proposed in Section IV is testified. Since it has been shown in the previous subsection that the proposed M-MMSE scheme performs better than the conventional techniques, especially for large $\beta$, we focus on the M-MMSE scheme in this subsection. Statistical channel inversion power control $p_{lk} = \frac{\rho}{d_l({\bf z}_{lk})}$ is still applied for pilots, while the uplink payload data power $\tau_{jk}$ is optimized. ${\rho}/{\sigma^2}$ is still set to 0 dB and the maximal transmit power $P_{max}$ in $\cal P$ is selected to make the cell edge SNR (without shadowing) equal to -3 dB. Results for the equal power allocation (i.e., $\tau_{lk} = P_{max}$) is provided as a base line. We also apply Algorithm~\ref{alg:alg1} to the instantaneous SINR in~(\ref{sinr_ul}) for comparison. The following results are obtained for $M = 100$ and $K = 10$. After generating user locations and shadow fading, 9 users with the worst channel conditions in the whole network are dropped to provide 95\% coverage.

\begin{figure}[tbp]
\vspace{-1ex}
\begin{minipage}[t]{0.46\linewidth}
\centering
\scalebox{0.48}{\includegraphics{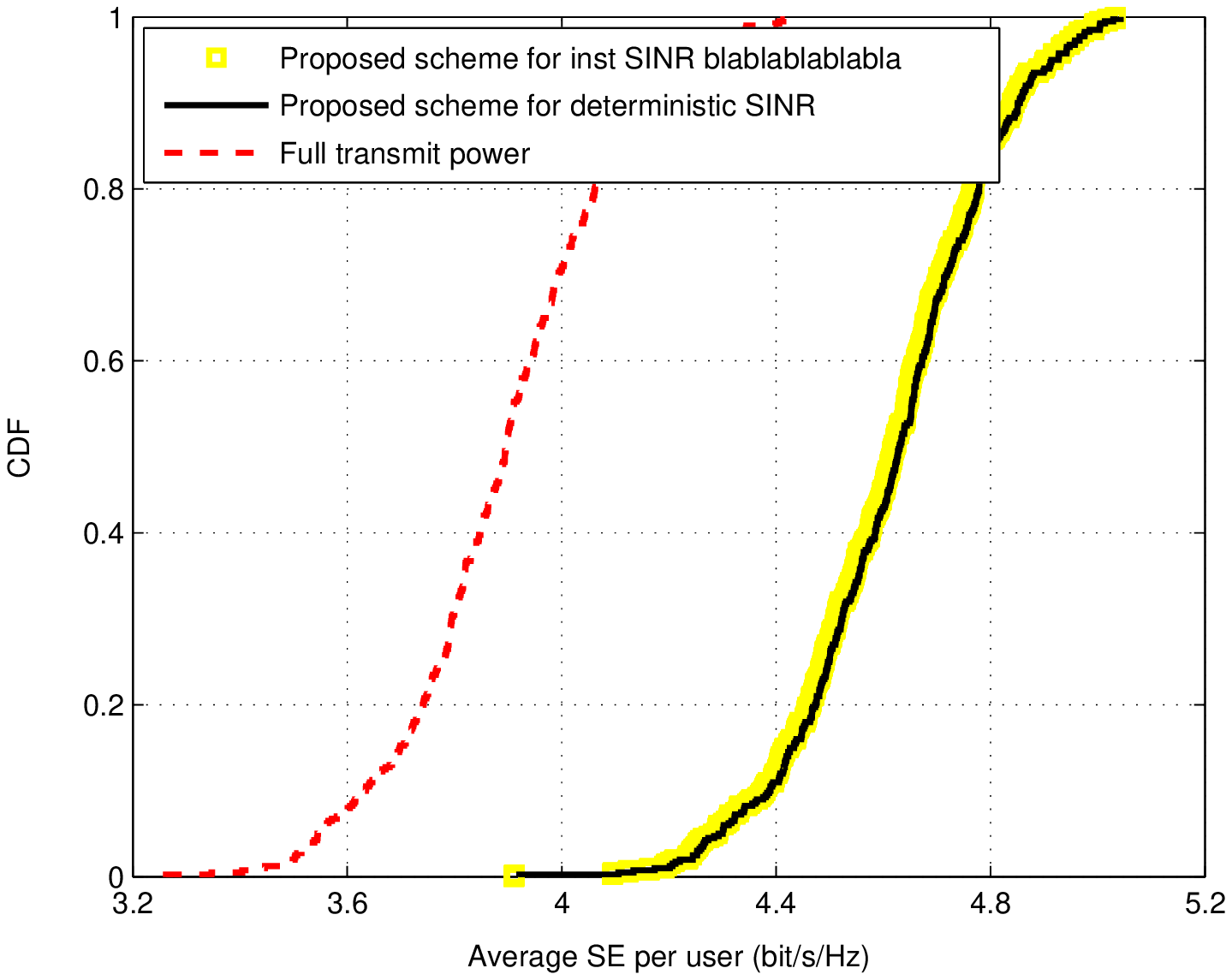}}
\caption{CDFs of average user SE with $\beta=4$.}
\label{avg_ue_rate_bete4}
\end{minipage}
\hfil
\begin{minipage}[t]{0.46\linewidth}
\centering
\scalebox{0.48}{\includegraphics{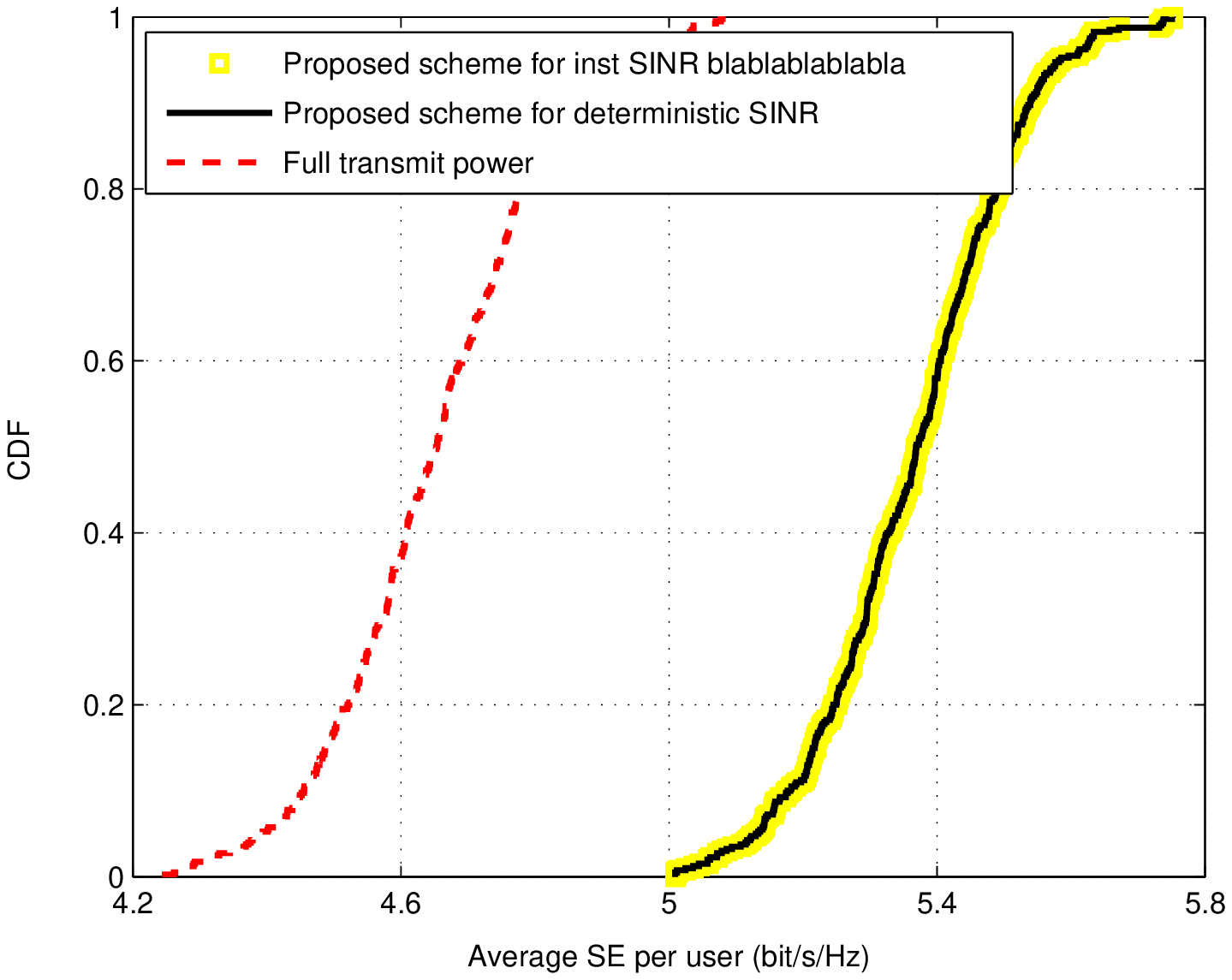}}
\caption{CDFs of average user SE with $\beta=7$.}
\label{avg_ue_rate_bete7}
\end{minipage}
\vspace{-3ex}
\end{figure}

We first consider the performance metric of average user SE which is calculated as the network sum SE divided by the number of served users. The cumulative distribution functions (CDFs) over user locations are shown in Fig.~\ref{avg_ue_rate_bete4} and Fig.~\ref{avg_ue_rate_bete7} for $\beta = 4$ and $\beta = 7$, respectively. As seen from the figures, the CDF curves with long-term power control based on Algorithm~\ref{alg:alg1} coincide with those with short-term power control optimized for the instantaneous SINR at every coherence block, which validates our power control based on the large-scale SINR approximation. Since the approximation only depends on the long-term statistics, the optimization complexity can be spread over time. Furthermore, compared with the equal power allocation policy, the average user SEs can be significantly improved by our power control scheme. At the 50 percentile, 17\% increase can be achieved by our scheme for both $\beta = 4$ and $\beta = 7$.

\psfrag{Per user rate (bit/s/Hz)}[][cb]{\Large {Per user SE (bit/s/Hz)}}
\psfrag{Average SE per user (bit/s/Hz)}[][cb]{\Large {Average user SE (bit/s/Hz)}}
\psfrag{CDF}[][]{\Large{CDF}}
\psfrag{Proposed scheme for inst SINR blablablablabla}{\Large {Proposed scheme for instant. SINR}}
\psfrag{Proposed scheme for deterministic SINR}{\Large {Proposed scheme for determ. SINR}}
\psfrag{Full transmit power}{\Large {Equal power allocation}}
\psfrag{Intra-cell equal throughput power control}{\Large {Intra-cell equal SE power control}}

\psfrag{0}[][l]{\Large {0}}
\psfrag{0.2}[][l]{\Large {0.2}}
\psfrag{0.4}[][l]{\Large {0.4}}
\psfrag{0.6}[][l]{\Large {0.6}}
\psfrag{0.8}[][l]{\Large {0.8}}
\psfrag{1}[][l]{\Large {1}}

\psfrag{2}[][]{\Large {2}}
\psfrag{4}[][]{\Large {4}}
\psfrag{6}[][]{\Large {6}}
\psfrag{8}[][]{\Large {8}}
\psfrag{10}[][]{\Large {10}}
\psfrag{12}[][]{\Large {12}}
\psfrag{14}[][]{\Large {14}}
\psfrag{16}[][]{\Large {16}}
\psfrag{18}[][]{\Large {18}}
\psfrag{20}[][]{\Large {20}}
\psfrag{22}[][]{\Large {22}}

\psfrag{3.2}[][]{\Large {3.2}}
\psfrag{3.6}[][]{\Large {3.6}}
\psfrag{4}[][]{\Large {4}}
\psfrag{4.4}[][]{\Large {4.4}}
\psfrag{4.8}[][]{\Large {4.8}}
\psfrag{5.2}[][]{\Large {5.2}}
\psfrag{4.2}[][]{\Large {4.2}}
\psfrag{4.6}[][]{\Large {4.6}}
\psfrag{5}[][]{\Large {5}}
\psfrag{5.4}[][]{\Large {5.4}}
\psfrag{5.8}[][]{\Large {5.8}}

\begin{figure}[tbp]
\begin{minipage}[t]{0.46\linewidth}
\centering
\scalebox{0.48}{\includegraphics{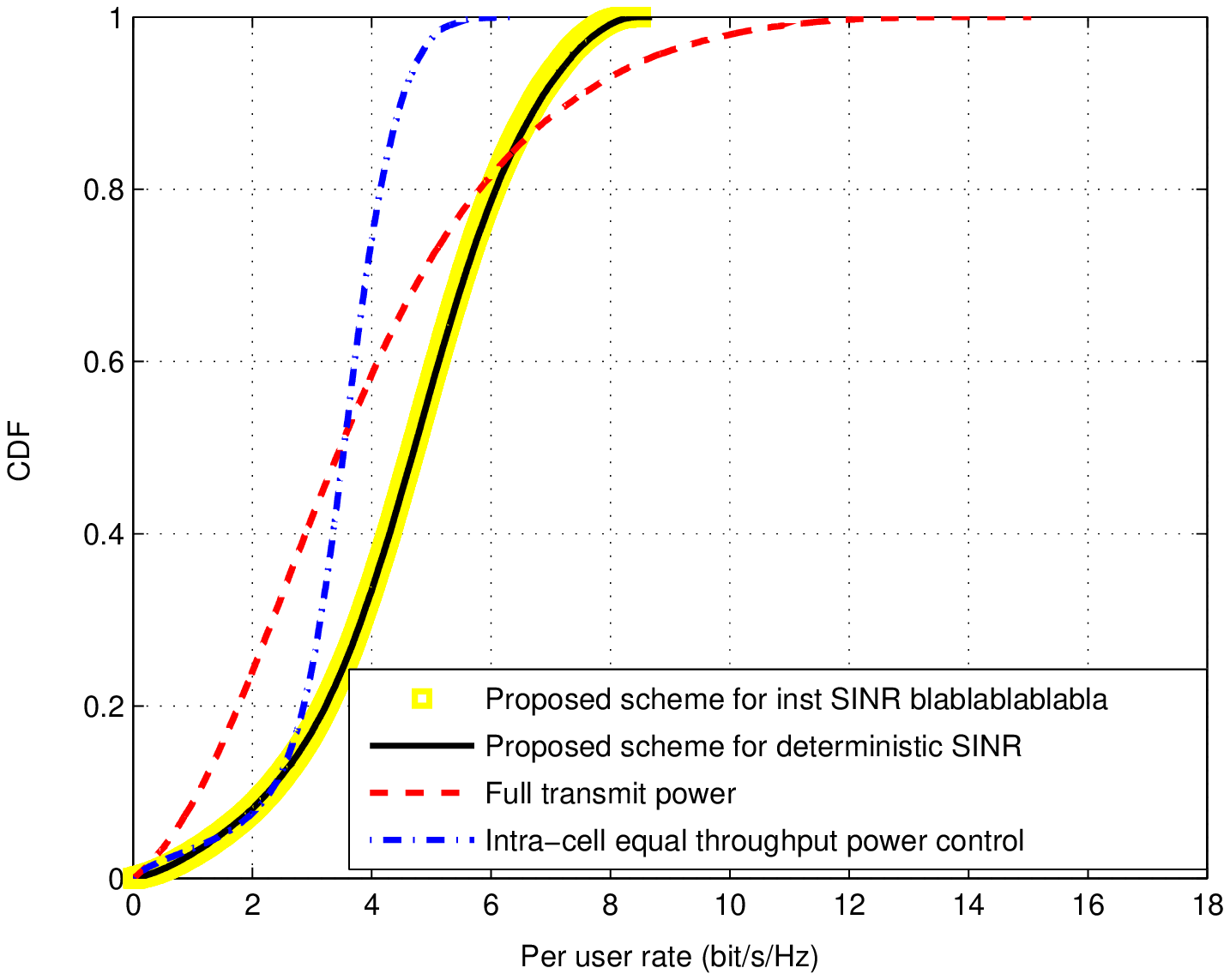}}
\caption{CDFs of per user SE with $\beta=4$.}
\label{per_ue_rate_bete4}
\end{minipage}
\hfil
\begin{minipage}[t]{0.46\linewidth}
\centering
\scalebox{0.48}{\includegraphics{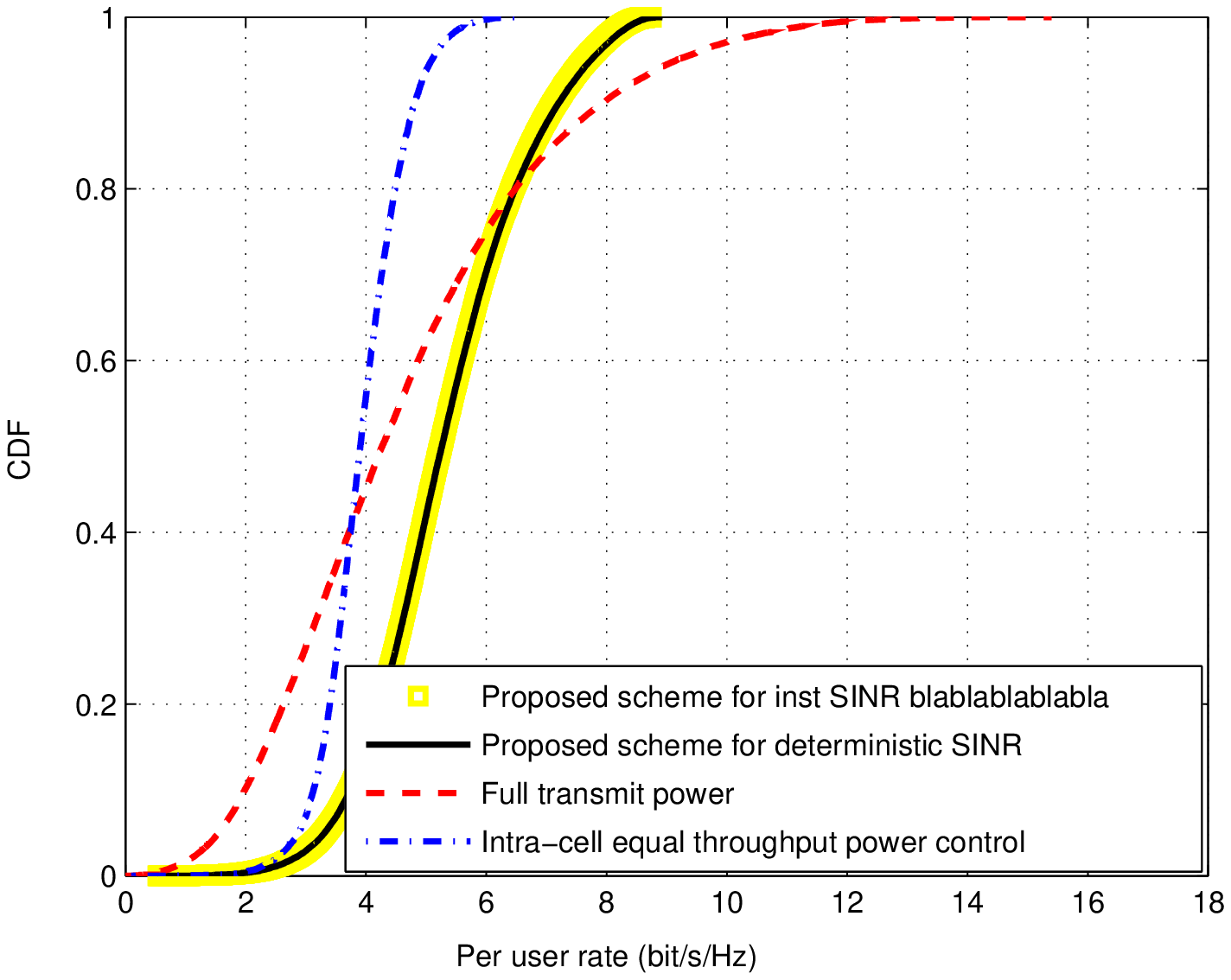}}
\caption{CDFs of per user SE with $\beta=7$.}
\label{per_ue_rate_bete7}
\end{minipage}
\vspace{-2ex}
\end{figure}

We analyze how the per user SE at different parts of the cells is affected by our power control. Results are also provided for the power control proposed in~\cite{Yang2014}, which tries to provide equal SE for users in the same cell so that, to some extent, intra-cell user fairness is guaranteed. CDFs of the per user SE are shown in Fig.~\ref{per_ue_rate_bete4} for $\beta = 4$ and in Fig.~\ref{per_ue_rate_bete7} for $\beta = 7$. Equal power allocation leads to the largest SE variations, while the power control from~\cite{Yang2014} gives relatively small variations. Interestingly, the proposed power control from Algorithm~\ref{alg:alg1} provides essentially the same SE for the weakest users, while pushing the SE of the majority of the users to higher values. Despite the larger SE variations, we conclude the proposed power control brings a better type of user fairness than the scheme from~\cite{Yang2014} since the strong users get higher SEs without degrading for the weakest ones.

\section{Conclusions} \label{conclusions}
In this paper, a new state-of-the-art multi-cell MMSE scheme is proposed, which includes an uplink M-MMSE detector and a downlink M-MMSE precoder. Compared with the conventional single-cell MMSE scheme, that only makes use of the intra-cell channel directions, the novelty of our multi-cell MMSE scheme is that it utilizes all channel directions that can be estimated locally at each BS, so that both intra-cell and inter-cell interference can be actively suppressed. The proposed scheme brings very promising sum SE gains over the conventional single-cell MMSE and the multi-cell ZF from~\cite{bjornson2014massive}, particularly for large $\beta$ and $K$. Since imperfect estimated CSI is accounted for in our scheme, the gains obtained by our scheme are likely to be achievable in practical systems. Furthermore, large-scale approximations of the uplink and downlink SINRs are derived for the proposed multi-cell MMSE scheme, and these are tight in the large-system limit. The approximations are easy to compute since they only depend on large-scale fading, power control and pilot allocation, and shown to be very accurate even for small system dimensions. Based on the SINR approximations, an uplink-downlink duality is established and a low complexity power control algorithm for sum SE maximization is proposed for the multi-cell MMSE scheme. The proposed power control brings a notable sum SE gain and also provides good user fairness compared to the equal power allocation policy. Since the SINR approximations depend only on long-term statistics, the complexity of the power control algorithm can be spread over a long time period.

\appendices
\section{Useful Lemmas} \label{lemmas}
\begin{Lemma} (Matrix inversion lemma (\Rmnum{1}), \cite{silverstein1995empirical}):\label{lemma1}
Let ${\bf A} \in \Cset^{M \times M}$ be a Hermitian invertible matrix. Then, for any vector ${\bf x} \in \Cset ^{M \times 1}$ and any scalar $\tau \in \Cset$ such that ${\bf A}+\tau{\bf{xx}}^H$ is invertible,
\begin{equation}
{\bf x}^H \left( {\bf A}+\tau{\bf{xx}}^H \right)^{-1} = \frac{{\bf x}^H {\bf A}^{-1}}{1+\tau {\bf x}^H {\bf A}^{-1}{\bf x}}.
\end{equation}
\end{Lemma}
\begin{Lemma}(Matrix inversion lemma (\Rmnum{2}),\cite{Hoydis2013}):\label{lemma2}
Let ${\bf A} \in \Cset^{M \times M}$ be a Hermitian invertible matrix. Then, for any vector ${\bf x} \in \Cset ^{M \times 1}$ and any scalar $\tau \in \Cset$ such that ${\bf A}+\tau{\bf{xx}}^H$ is invertible,
\begin{equation}
\left( {\bf A}+\tau{\bf{xx}}^H \right)^{-1} = {\bf A}^{-1} - \frac{\tau {\bf A}^{-1}{\bf x}{\bf x}^H {\bf A}^{-1}}{1+\tau {\bf x}^H {\bf A}^{-1}{\bf x}}.
\end{equation}
\end{Lemma}
\begin{Lemma}(Generalized rank-1 perturbation lemma, Lem. 14.3 in~\cite{couillet2011random}):\label{lemma3} % lemma 14.3 in page 352 of random matrix methods for wireless communications
Let ${\bf A} \in \Cset^{M \times M}$ be deterministic with uniformly bounded spectral norm (with respect to $M$) and ${\bf B} \in \Cset^{M \times M}$ be a random Hermitian matrix, with eigenvalues $\lambda_{1}^{\bf B} \le ... \le \lambda_{M}^{\bf B}$ such that, with probability one, there exist $\varepsilon >0$ and $M_0$ such that $\lambda_{1}^{\bf B} > \varepsilon$ for all $M > M_0$. Then for any vector ${\bf v} \in \Cset^{M \times 1}$,
\begin{equation}
\frac{1}{M}{\rm{tr}}\left({\bf A}{\bf B}^{-1}\right) - \frac{1}{M}{\rm{tr}}\left({\bf A}{\left({\bf B} +{\bf {vv}}^H \right)^{-1}}\right)  \xrightarrow[M \to \infty]{a.s} 0
%| {\rm{tr}} \left( \left( {\bf B} - z{\bf I}_M \right)^{-1} - \left( {\bf B} +{\bf {vv}}^H - z{\bf I}_M \right)^{-1} \right){\bf A}| \le \frac{\left\| {\bf A} \right\|}{|z|}.
\end{equation}
where ${\bf B}^{-1}$ and $\left({\bf B} +{\bf {vv}}^H \right)^{-1}$ exist with probability one.
\end{Lemma}
\begin{Lemma}(Lem. B.26 in \cite{bai2009spectral}, Thm. 3.7 in \cite{couillet2011random}, Lem. 12 in \cite{Hoydis2012thesis}):\label{lemma4}
Let ${\bf A} \in \Cset^{M \times M}$ and ${\bf x}$, ${\bf y} \sim {\cal {CN}} (0, \frac{1}{M}{\bf I}_M )$. Assume that ${\bf A}$ has uniformly bounded spectral norm (with respect to $M$) and that ${\bf x}$, ${\bf y}$ and ${\bf A}$ are mutually independent. Then, for all $p \ge 1$,
%\begin{center}
\begin{enumerate}
\item  $ \Eset \left\{\left|{\bf x}^H {\bf A} {\bf x} - \frac{1}{M}{\rm{tr}}\left({\bf A}\right)\right|^p \right\} ={\cal O}\left( \frac{1}{{M}^{\frac{p}{2}}}\right) $
\item  ${\bf x}^H {\bf A} {\bf x}-\frac{1}{M}{\rm{tr}}\left({\bf A}\right) \xrightarrow[M \to \infty]{a.s} 0$
\item  ${\bf x}^H {\bf A} {\bf y} \xrightarrow[M \to \infty]{a.s} 0$
\item  $\Eset \left\{\left|\left({\bf x}^H {\bf A} {\bf x}\right)^2 - \left(\frac{1}{M}{\rm{tr}}\left({\bf A}\right)\right)^2\right| \right\} \xrightarrow[M \to \infty]{} 0$.
\end{enumerate}
%\end{center}
\end{Lemma}

\section{Proof of Theorem~\ref{theorem3}}\label{sec:proof_thr3}
Define ${\bf \Sigma}_j = ({{\hat{\bf H}}_{{\mathcal{V}},{j}}}{{\bf{\Lambda }}_j}{{\hat{\bf H}}_{{\mathcal{V}},{j}}}^{H} + \left(\sigma^2+ \varphi_j \right){\bf I}_M)^{-1}$, then the M-MMSE detector in~\ref{detector2} is ${{\bf{g}}_{jk}} = {\bf \Sigma}_j{\hat {\bf h}}_{jjk}$. We omit the superscript ``M-MMSE'' in the proof for brevity. In the following proof, we use $\asymp$ to denote the almost sure convergence such that $a \asymp b$ represents $a - b \xrightarrow[M\to \infty]{a.s.}0$. Define
\begin{enumerate}
\item ${{\hat {\bf H}}_{{\mathcal{V}},{jlk}}} = \left[ { \hat{\bf h}}_{{\cal V},j1},...,{\hat {\bf h}}_{{\cal V},j\left( i_{lk} -1 \right)},{\hat {\bf h}}_{{\cal V},j\left( i_{lk} + 1 \right)},..., {\hat{\bf h}}_{{\cal V},jB} \right]$,
\item ${\bf \Lambda}_{jlk}={\rm{diag}}\left( \lambda_{j1},...\lambda_{j\left( i_{lk} - 1 \right)},\lambda_{j\left( i_{lk} + 1 \right)},....,\lambda_{jB}\right)$,
\item ${\bf \Sigma}_{jjk}= \left({{\hat {\bf H}}_{{\mathcal{V}},{jjk}}}{{\bf{\Lambda }}_{jjk}}{{\hat{\bf H}}_{{\mathcal{V}},{jjk}}}^{H} + \left(\sigma^2+ \varphi_j \right){\bf I}_M\right)^{-1}$,
\item ${\bf \Sigma}_j^{'} = M{\bf \Sigma}_j$ and ${\bf \Sigma}_{jjk}^{'} = M {\bf \Sigma}_{jjk}$,
\end{enumerate}
then we have the following lemma.
\begin{Lemma} \label{lemma5}
Let ${{\hat {\bf h}}_{jlk}}$ and ${{\tilde{\bf h}}_{jlk}}$ denote the MMSE estimate of ${\bf h}_{jlk}$ as in~(\ref{estimation2}) and its estimation error, respectively, then
\begin{equation}
{{ \hat {\bf h}}_{jjk}^H} {\bf \Sigma}_{j} {\hat{\bf h}}_{jjk}- \frac{p_{jk} d_j^2\left({\bf z}_{jk} \right)
\delta_{jk}}{1+\lambda_{ji_{jk}}\delta_{jk}} \xrightarrow[M \to \infty]{a.s.} 0,
\end{equation}
\begin{equation}
{ \hat {\bf h}}_{jjk}^H {\bf \Sigma}_{j} {\tilde{\bf h}}_{jlm} \xrightarrow[M \to \infty]{a.s.} 0.
\end{equation}
\end{Lemma}
\noindent\emph{Proof:}
Let $ x = {\hat {\bf h}}_{jjk}^H {\bf \Sigma}_{j}{\hat{\bf h}}_{jjk} $, then
\begin{eqnarray} \label{x}
x &= &{\hat {\bf h}}_{jjk}^H \left({\bf \Sigma}_{jjk}^{-1} + \lambda_{ji_{jk}}{ \hat{\bf h}}_{{\cal V},ji_{jk}}{\hat{\bf h}}_{{\cal V},ji_{jk}}^H \right)^{-1}{\hat{\bf h}}_{jjk} \nonumber \\
&\mathop  = \limits^{\left( a \right)} & \frac{ p_{jk} d_j^2\left({\bf z}_{jk} \right) {\hat {\bf h}}_{{\cal V},ji_{jk}}^H {\bf \Sigma}_{jjk} {\hat {\bf h}}_{{\cal V},ji_{jk}} }{1+\lambda_{ji_{jk}}
{\hat {\bf h}}_{{\cal V},ji_{jk}}^H {\bf \Sigma}_{jjk} {\hat {\bf h}}_{{\cal V},ji_{jk}} }
= \frac{\frac{1}{M}p_{jk} d_j^2\left({\bf z}_{jk} \right) {\hat {\bf h}}_{{\cal V},ji_{jk}}^H {\bf \Sigma}_{jjk}^{'}{\hat {\bf h}}_{{\cal V},ji_{jk}}}{1+\lambda_{ji_{jk}}\frac{1}{M}{\hat {\bf h}}_{{\cal V},ji_{jk}}^H {\bf \Sigma}_{jjk}^{'} {\hat {\bf h}}_{{\cal V},ji_{jk}} } \nonumber \\
%\mathop  \asymp \limits^{\left( b \right)} \frac{{\rm{tr}} \left({\bf \Phi}_{jjk} {\bf \Sigma}_{jjk}\right)}{1 + \lambda_{ji_{jk}} {\rm{tr}}\left({\bf \Phi}_{jjk}{\bf \Sigma}_{jjk} \right)}
&\mathop  \asymp \limits^{\left( b \right)}&\frac{\frac{1}{M} p_{jk} d_j^2\left({\bf z}_{jk} \right) {\rm{tr}} \left(
{\tilde{\bf \Phi}}_{{\cal V},ji_{jk}} {\bf \Sigma}_{jjk}^{'}\right)}{1 + \frac{1}{M} \lambda_{ji_{jk}} {\rm{tr}}\left(
{\tilde{\bf \Phi}}_{{\cal V},ji_{jk}}{\bf \Sigma}_{jjk}^{'} \right)}
\mathop \asymp \limits^{\left( c\right)} \frac{\frac{1}{M}p_{jk} d_j^2\left({\bf z}_{jk} \right)  {\rm{tr}} \left({\tilde{\bf \Phi}}_{{\cal V},ji_{jk}} {\bf \Sigma}_{j}^{'}\right)}{1 + \frac{1}{M} \lambda_{ji_{jk}} {\rm{tr}}\left({\tilde{\bf \Phi}}_{{\cal V},ji_{jk}}{\bf \Sigma}_{j}^{'} \right)}\nonumber \\
&\mathop  \asymp \limits^{\left( d \right)}& \frac{ \frac{1}{M}p_{jk} d_j^2\left({\bf z}_{jk} \right)  {\rm{tr}} \left({\tilde{\bf \Phi}}_{{\cal V},ji_{jk}} {\bf T}_j \right) }{1 + \lambda_{ji_{jk}}  \frac{1}{M} {\rm{tr}} \left({\tilde{\bf \Phi}}_{{\cal V},ji_{jk}} {\bf T}_j \right) }
\mathop = \limits^{\left( e \right)} \frac{ p_{jk} d_j^2\left({\bf z}_{jk} \right)\delta_{jk}}{1+\lambda_{ji_{jk}}\delta_{jk}},
\end{eqnarray}
where $(a)$ follows from Lemma~\ref{lemma1} and the fact that ${\hat{\bf h}}_{jjk} = \sqrt{p_{jk}} d_j({\bf z}_{jk})
{\hat {\bf h}}_{{\cal V},ji_{jk}} $ and $( b )$ follows from Lemma~\ref{lemma4} $2)$. Notice that Lemma~\ref{lemma4} $2 )$ can be applied since ${\bf \Sigma}_{jjk}^{'} =(\frac{1}{M} {\hat{\bf H}}_{{\cal V},jjk}{\bf \Lambda}_{jjk}{\hat{\bf H}}_{{\cal V},jjk}^H + \frac{\sigma^2+ \varphi_j}{M} {\bf I}_M )^{-1}$ has uniformly bounded spectral norm as $M \to \infty$, because $\varphi_j$ scales as $K$ and $\frac{K}{M} > 0$ by assumption, thus $\frac{\varphi_j}{M} > 0$ for all $M$. $(c )$ follows from Lemma~\ref{lemma3}. $(d)$ follows from Theorem~\ref{theorem1} for ${\bf D}={\tilde{\bf \Phi}}_{{\cal V},ji_{jk}}$, ${\bf T}_j = {\bf T}_j(\frac{\sigma^2 +  \varphi_j}{M})$ with ${\bf R}_b =\lambda_{jb}{\tilde{\bf \Phi}}_{{\cal{V}},jb}$. In step $( e)$, we use the notation $\delta_{jk} = \frac{1}{M} {\rm{tr}}({\tilde{\bf \Phi}}_{{\cal V},ji_{jk}} {\bf T}_j)$ and arrive at the expression in (42).

Let $ y = {\hat{\bf h}}_{jjk}^H {\bf \Sigma}_{j} {\tilde{\bf h}}_{jlm} $, then
\begin{eqnarray}
y &= &{\hat {\bf h}}_{jjk}^H \left({\bf \Sigma}_{jjk}^{-1} + \lambda_{ji_{jk}}{ \hat{\bf h}}_{{\cal V},ji_{jk}}{\hat{\bf h}}_{{\cal V},ji_{jk}}^H \right)^{-1}{\tilde{\bf h}}_{jlm} \nonumber \\
&\mathop  = \limits^{\left( a \right)} & \frac{\sqrt{p_{jk}} d_j\left({\bf z}_{jk} \right) {\hat {\bf h}}_{{\cal V},ji_{jk}}^H {\bf \Sigma}_{jjk} {\tilde{\bf h}}_{jlm}}{1+\lambda_{ji_{jk}}{\hat {\bf h}}_{{\cal V},ji_{jk}}^H {\bf \Sigma}_{jjk} {\hat {\bf h}}_{{\cal V},ji_{jk}} }
=\frac{\sqrt{p_{jk}} d_j\left({\bf z}_{jk} \right)\frac{1}{M}{\hat {\bf h}}_{{\cal V},ji_{jk}}^H {\bf \Sigma}_{jjk}^{'} {\tilde{\bf h}}_{jlm} }{1+\lambda_{ji_{jk}}\frac{1}{M}{\hat {\bf h}}_{{\cal V},ji_{jk}}^H {\bf \Sigma}_{jjk}^{'} {\hat {\bf h}}_{{\cal V},ji_{jk}} } \mathop \asymp \limits^{\left( b \right)}0,
\end{eqnarray}
where steps $( a )$ and $b$ follow from Lemma~\ref{lemma1} and Lemma~\ref{lemma4}~$3)$, respectively, which completes the proof.  \hfill{$\blacksquare$}

We use this lemma in the following to determine the asymptotic behaviour of each term in the uplink SINR of~(\ref{sinr_ul}).
\vspace{-1ex}
\subsection{Signal power}
Since ${\bf g}_{jk}^H {\hat{\bf h}_{jjk}} =
{\hat {\bf h}}_{jjk}^H {\bf \Sigma}_{j} {\hat{\bf h}_{jjk}} $, then according to Lemma~\ref{lemma5}, it is obvious that
\begin{equation}
{\bf g}_{jk}^H {\hat{\bf h}_{jjk}} - \frac{p_{jk} d_j^2\left({\bf z}_{jk} \right) \delta_{jk}}{1+\lambda_{ji_{jk}}\delta_{jk}} \xrightarrow[M \to \infty]{a.s.} 0.
\end{equation}
By the continuous mapping theorem~\cite{Vaart2000}, we further obtain
\begin{equation}
\left|{\bf g}_{jk}^{H} {\hat{\bf h}}_{jjk} \right|^2 - \left( \frac{p_{jk} d_j^2\left({\bf z}_{jk} \right) \delta_{jk}}{1+\lambda_{ji_{jk}}\delta_{jk}} \right)^2  \xrightarrow[M \to \infty]{a.s.} 0.
\end{equation}
\vspace{-1ex}
\subsection{Channel uncertainty}
According to Lemma~\ref{lemma5},
\begin{equation}
{\bf g}_{jk}^H {\tilde{\bf h}_{jjk}} = {\hat{\bf h}_{jjk}}^H {\bf \Sigma}_j {\tilde{\bf h}_{jjk}} \xrightarrow[M \to \infty]{a.s.}0.
\end{equation}
Thus by the dominated convergence theorem~\cite{Billingsley1995} and the continuous mapping theorem, we have
\begin{equation}
\Eset\left\{\tau_{jk}\left|{\bf g}_{jk}^H {\tilde{\bf h}_{jjk}}\right|^2 \bigg | {\hat{\bf h}}_{(j)} \right\}  \xrightarrow[M \to \infty]{a.s.}0.
\end{equation}
\vspace{-5ex}
\subsection{Interference power}
Since ${\bf{g}}_{jk}^H={\bf \Sigma}_j {\hat {\bf h}}_{jjk}$, the interference power from user $m$ in cell $l$ is ${\Eset_{\{ {\bf{h}} \}}}\{ | {{\bf{g}}_{jk}^H{{\bf{h}}_{jlm}}}|^2  | {\hat{\bf h}}_{(j)}\} = \Eset \{ | {\hat {\bf h}}_{jjk}^{H} {\bf \Sigma}_j {\bf  h}_{jlm} |^2 | {\hat{\bf h}}_{(j)}\}$. The computation depends on which pilots that are used.

\subsubsection{$i_{lm} = i_{jk} = i_0$}
In this case, user $k$ in cell $j$ use the same pilot sequence as user $m$ in cell $j$, and there will be coherence pilot contaminated interference. Since
\begin{equation}
{{\hat {\bf {h}}}_{jlm}} = \sqrt{p_{lm}} d_j\left({\bf z}_{lm} \right)  {\hat{\bf h}}_{{\cal V},ji_0} = \sqrt{\frac{p_{lm}}{p_{jk}}} \frac{d_j\left({\bf z}_{lm}\right) }{d_j\left({\bf z}_{jk} \right)}{{\hat{\bf {h}}}_{jjk}},
\end{equation}
we have
\begin{eqnarray}
{\hat {\bf h}}_{jjk}^H{\bf \Sigma}_j {{\bf h}}_{jlm}= \sqrt{\frac{p_{lm}}{p_{jk}}} \frac{d_j\left({\bf z}_{lm}\right) }{d_j\left({\bf z}_{jk} \right)} {\hat {\bf h}}_{jjk}^H {\bf \Sigma}_j {\hat {\bf h}}_{jjk} + {\hat {\bf h}}_{jjk}^H{\bf \Sigma}_j{\tilde {\bf h}}_{jlm}
\mathop \asymp \limits^{\left( a \right)} d_j\left({\bf z}_{jk} \right) d_j\left({\bf z}_{lm} \right)  \frac{\sqrt{p_{jk}p_{lm}}\delta_{jk}}{1+\lambda_{ji_{jk}}\delta_{jk}},
\end{eqnarray}
where in step $ (a)$ the first term remains and the second term vanishes according to Lemma~\ref{lemma5}. Indicated by the dominated convergence theorem and the continuous mapping theorem, we have
\begin{equation}
\Eset \left\{ \left| { \hat {\bf h}}_{jjk}^{H} {\bf \Sigma}_j {\bf  h}_{jlm} \right|^2\bigg | {\hat{\bf h}}_{(j)} \right\} - {d_j^2\left({\bf z}_{jk} \right)}d_j^2\left({\bf z}_{lm} \right)  \frac{{p_{jk}p_{lm}}\delta_{jk}^2}{\left(1+\lambda_{ji_{jk}}\delta_{jk}\right)^2} \xrightarrow[M \to \infty]{a.s.}0.
\end{equation}
\vspace{-1ex}
\subsubsection{$i_{lm} \ne i_{jk}$}
In this case, two users have different pilots, such that
\begin{eqnarray} \label{temp3}
\left| {\hat {\bf h}}_{jjk}^H{\bf \Sigma}_j {\bf  h}_{jlm} \right|^2 &\mathop =\limits^{\left( a \right)}& p_{jk} d_j^2\left({\bf z}_{jk} \right) \frac{ \frac{1}{M^2}{\hat {\bf h}}_{{\cal V},ji_{jk}}^H {\bf \Sigma}_{jjk}^{'} {\bf  h}_{jlm} {\bf h}_{jlm}^{H} {\bf \Sigma}_{jjk}^{'} {\hat {\bf h}}_{{\cal V},ji_{jk}} }{ \left( 1+\lambda_{ji_{jk}}{\hat {\bf h}}_{{\cal V},ji_{jk}}^H {\bf \Sigma}_{jjk} {\hat {\bf h}}_{{\cal V},ji_{jk}}\right)^2} \nonumber \\
&\mathop  \asymp \limits^{\left( b \right)}& p_{jk} d_j^2\left({\bf z}_{jk} \right) \frac{\frac{1}{M^2}{\rm{tr}}\left(
{\tilde{\bf \Phi}}_{{\cal V},ji_{jk}} {\bf \Sigma}_{jjk}^{'} {\bf  h}_{jlm} {\bf h}_{jlm}^{H} {\bf \Sigma}_{jjk}^{'} \right)}{\left(1+\lambda_{ji_{jk}}\delta_{jk} \right)^2}\nonumber \\
& =& p_{jk} d_j^2\left({\bf z}_{jk} \right) \frac{{\rm{tr}}\left( {\tilde{\bf \Phi}}_{{\cal V},ji_{jk}} {\bf \Sigma}_{jjk} {\bf  h}_{jlm} {\bf h}_{jlm}^{H} {\bf \Sigma}_{jjk} \right)}{\left(1+\lambda_{ji_{jk}}\delta_{jk} \right)^2}\nonumber \\
&=&p_{jk} d_j^2\left({\bf z}_{jk} \right) \frac{ {{\bf{h}}_{jlm}^H}{{\mathbf{\Sigma }}_{jjk}}{\tilde{\bf \Phi}}_{{\cal V},ji_{jk}} {{\bf{\Sigma }}_{jjk}}{\mathbf{h}}_{jlm}}{{\left( {1 + {\lambda _{j{i_{jk}}}}{\delta _{jk}}} \right)}^2},
\end{eqnarray}
where step $( a)$ follows from Lemma~\ref{lemma1} and the definition of ${\bf \Sigma}_{jjk}^{'}$. Step $( b )$ follows Lemma~\ref{lemma4}~$2 )$, Lemma~\ref{lemma3} and Theorem~\ref{theorem1}. It remains to obtain a deterministic equivalent of the numerator in~(\ref{temp3}). Define ${\bf \Sigma }_{j,jk,lm} = {( {{\mathbf{\Sigma }}_{jjk}^{ - 1} - {\lambda _{j{i_{lm}}}}{{{\hat{\bf h}}}_{{\cal V},j{i_{lm}}}}{\hat{\bf h}}_{{\cal V},j{i_{lm}}}^H } )^{ - 1}}$, then according to Lemma~\ref{lemma2} we have
\begin{equation} \label{temp2}
{{\mathbf{\Sigma }}_{jjk}} = {{\mathbf{\Sigma }}_{j,jk,lm}} - \frac{{{{\mathbf{\Sigma }}_{j,jk,lm}}{\lambda _{j{i_{lm}}}}{{{\hat{\bf h}}}_{{\cal V},j{i_{lm}}}}{\hat{\bf h}}_{{\cal V},j{i_{lm}}}^H{{\mathbf{\Sigma }}_{j,jk,lm}}}}{{1 + {\lambda _{j{i_{lm}}}}{\hat{\bf h}}_{{\cal V},j{i_{lm}}}^H{{\mathbf{\Sigma }}_{j,jk,lm}}{{{\hat{\bf h}}}_{{\cal V},j{i_{lm}}}}}}.
\end{equation}
Plugging~(\ref{temp2}) into the numerator of~(\ref{temp3}), we obtain
\begin{eqnarray} \label{temp4}
{{ {\bf h}}_{jlm}^H{{\mathbf{\Sigma }}_{jjk}}{\tilde{\bf \Phi}}_{{\cal V},ji_{jk}}{{\mathbf \Sigma }_{jjk}}{\hat {\bf h}}_{jlm}} =\underbrace { {\mathbf{h}}_{jlm}^H{{\mathbf{\Sigma }}_{j,jk,lm}}{\tilde{\bf \Phi}}_{{\cal V},ji_{jk}}{{\mathbf{\Sigma }}_{j,jk,lm}}{\mathbf{h}}_{jlm}}_{\left({\rm{intf.}}\,1 \right)} \nonumber \\
- 2\operatorname{Re} \left\{ {\underbrace {\frac{{\lambda _{j{i_{lm}}}} {\bf h}_{jlm}^H {{\mathbf{\Sigma }}_{j,jk,lm}}{\tilde{\bf \Phi}}_{{\cal V},ji_{jk}}{{\mathbf{\Sigma }}_{j,jk,lm}}{\hat {\bf h}}_{{\cal V},ji_{lm}}{\hat {\bf h}}_{{\cal V},ji_{lm}}^H{{\mathbf{\Sigma }}_{j,jk,lm}}{\mathbf{h}}_{jlm}}{1 + {\lambda _{j{i_{lm}}}}{\hat {\bf h}}_{{\cal V},ji_{lm}}^H{{\mathbf{\Sigma }}_{j,jk,lm}}{\hat {\bf h}}_{{\cal V},ji_{lm}}}}_{\left( {\rm{intf.}}\,2\right)}} \right\} \nonumber \\
+ {\left| {{\lambda _{j{i_{lm}}}}} \right|^2}\underbrace {\frac{{{\left| {\bf h}_{jlm}^H{\mathbf{\Sigma }}_{j,jk,lm}{\hat{\bf h}}_{{\cal V},j{i_{lm}}}^H \right|}^2}{\hat{\bf h}}_{{\cal V},j{i_{lm}}}^H{\bf{\Sigma}}_{j,jk,lm}{\tilde{\bf \Phi}}_{{\cal V},ji_{jk}}{{\bf{\Sigma}}_{j,jk,lm}}{\hat {\bf h}}_{{\cal V},j{i_{lm}}}}{{{\left| {1 + {\lambda _{j{i_{lm}}}}{\hat{\bf h}}_{{\cal V},ji_{lm}}^H{{\bf{\Sigma }}_{j,jk,lm}}{\hat{\bf h}}_{{\cal V},ji_{lm}}} \right|}^2}}}_{\left( {\rm{intf.}}\,3\right)}.
\end{eqnarray}

\textit{Deterministic equivalent of $( {\rm{intf.}}\,1  )$:}
Define ${{\mathbf{\Sigma }}_{j,jk,lm}^{'}} = M{{\mathbf{\Sigma }}_{j,jk,lm}}$, then following similar procedures as before, it is straightforward to show that
\begin{eqnarray}\label{a}
{{\bf h}}_{jlm}^H{\bf{\Sigma }}_{j,jk,lm}{\tilde{\bf \Phi}}_{{\cal V},ji_{jk}}{\bf{\Sigma }}_{j,jk,lm}{\bf{h}}_{jlm} & \asymp & \frac{ d_j\left( {\bf z}_{lm} \right)}{M^2}{\rm{tr}}\left( {\bf{\Sigma }}_{j,jk,lm}^{'}{\tilde{\bf \Phi}}_{{\cal V},ji_{jk}}{{\bf{\Sigma }}_{j,jk,lm}^{'}} \right) \nonumber \\
&\asymp&\frac{ d_j\left( {\bf z}_{lm} \right)}{M^2}{\rm{tr}}\left( {{\bf{\Sigma }}_j^{'}{\tilde{\bf \Phi}}_{{\cal V},ji_{jk}}{\bf{\Sigma }}_j^{'}} \right)
\asymp \frac{ d_j\left( {\bf z}_{lm} \right)}{M^2}{\text{tr}}\left( {{{{\bf{T}}}_{jk}^{'}}} \right),
\end{eqnarray}
where ${\bf{T}}_{jk}^{'} = {\bf{T}}_{jk}^{'}( \alpha )$ is given by Theorem~\ref{theorem2} for $\alpha  = \frac{{{\sigma ^2} + {\varphi_j}}}{M}$ , ${\bf D} = {\bf I}_M$, ${\bf \Theta} = {\tilde{\bf \Phi}}_{{\cal V},ji_{jk}}$ and ${\bf R}_b =\lambda_{jb}{\bf \tilde \Phi}_{{\cal{V}},jb}$. %Thus,

\textit{Deterministic equivalent of $( {\rm{intf.}}\,2 )$:}
Instead of tackling the expression in $({\rm{intf.}}\,2)$ directly, we derive the deterministic equivalents of its numerator and denominator, respectively. Plugging ${\bf h}_{jlm} = {\hat {\bf h}}_{jlm} + {\tilde {\bf h}}_{jlm}$ and ${\hat {\bf h}}_{jlm}=\sqrt{p_{lm}} d_j({\bf z}_{lm}){\hat{\bf h}}_{{\cal V},ji_{lm}}$ into the numerator, we have that
\begin{eqnarray} \label{intf21}
{\bf h}_{jlm}^H{\bf{\Sigma }}_{j,jk,lm}{\tilde{\bf \Phi}}_{{\cal V},ji_{jk}}{\bf{\Sigma}}_{j,jk,lm}{{{\hat{\bf h}}}_{{\cal V},j{i_{lm}}}} &\mathop  \asymp \limits^{\left( a \right)}& \frac{\sqrt{p_{lm}} d_j\left({\bf z}_{lm} \right) }{M^2}{\rm{tr}}\left({\tilde{\bf \Phi}}_{{\cal V},ji_{lm}}{\bf{\Sigma}}_j^{'}{\tilde{\bf \Phi}}_{{\cal V},ji_{jk}}{{\bf{\Sigma }}_j^{'}} \right) \nonumber \\
&\mathop  \asymp \limits^{\left( b \right)}& \frac{\sqrt{p_{lm}} d_j\left({\bf z}_{lm} \right) }{M^2}{\text{tr}}\left( {{\tilde{\bf \Phi}}_{{\cal V},ji_{lm}}{\bf{T}}_{jk}^{'}} \right)\nonumber \\
&=& \sqrt{p_{lm}} d_j\left({\bf z}_{lm} \right)\frac{\vartheta _{jlmk}^{'}}{M}
\end{eqnarray}
by defining ${\vartheta}_{jlmk}^{'} = \frac{1}{M} {\rm{tr}} ({\tilde{\bf \Phi}}_{{\cal V},ji_{lm}} {\bf T}_{jk}^{'})\in \Rset$. Step $(a)$ follows from Lemma~\ref{lemma4} $2)$ and Lemma~\ref{lemma3}. Step $(b)$ follows from Theorem~\ref{theorem2}. Similarly, we have
\begin{eqnarray} \label{intf22}
{\hat{\bf h}}_{{\cal V},j{i_{lm}}}^H{{\bf{\Sigma }}_{j,jk,lm}}{\bf{h}}_{jlm} &\asymp& \frac{\sqrt{p_{lm}} d_j\left({\bf z}_{lm} \right) }{M}{\text{tr}}\left( {\tilde{\bf \Phi}}_{{\cal V},ji_{lm}}{\bf{\Sigma }}_j^{'} \right) \nonumber \\
&\asymp& \frac{\sqrt{p_{lm}} d_j\left({\bf z}_{lm} \right) }{M}{\text{tr}}\left( {\tilde{\bf \Phi}}_{{\cal V},ji_{lm}}{{\bf{T}}_j} \right) = \sqrt{p_{lm}} d_j\left({\bf z}_{lm} \right) \vartheta _{jlm},
\end{eqnarray}
\begin{equation}\label{intf23}
{\hat{\bf h}}_{{\cal V},j{i_{lm}}}^H{{\bf{\Sigma }}_{j,jk,lm}}{\hat{\bf h}}_{{\cal V},j{i_{lm}}}^H \asymp \vartheta _{jlm},
\end{equation}
where $\vartheta _{jlm}$ is defined as $\vartheta _{jlm} = \frac{1}{M} {\rm{tr}} ({\tilde{\bf \Phi}}_{{\cal V},ji_{lm}} {\bf T}_{j}) \in \Rset$ and ${\mathbf{T}}_j$ is given in Lemma~\ref{lemma5}. Based on~(\ref{intf21}) --~(\ref{intf23}), the equivalents of the denominator and numerator are given as $1 + \lambda _{ji_{lm}}\vartheta _{jlm}$ and $\frac{1}{M}\vartheta _{jlmk}^{'}\vartheta _{jlm}p_{lm} d_j^2({\bf z}_{lm} ) \lambda _{ji_{lm}}$, respectively. Therefore, according to the continuous mapping theorem, we have
\begin{equation} \label{b}
\left( {\rm{intf.}}\,2 \right) - \frac{\vartheta _{jlmk}^{'}\vartheta _{jlm}}{1 + \lambda _{ji_{lm}}\vartheta _{jlm}} \frac{p_{lm} d_j^2\left({\bf z}_{lm} \right) \lambda _{ji_{lm}}}{M} \xrightarrow[M \to \infty]{a.s.}0.
\end{equation}

\textit{Deterministic equivalent of $( {\rm{intf.}}\,3)$:}
Based on the techniques used to characterize $( {\rm{intf.}}\,1  )$ and $({\rm{intf.}}\,2 )$, it is straightforward to show that
\begin{equation} \label{c}
\left( {\rm{intf.}}\,3 \right) - \frac{{{{\left| {{\vartheta _{jlm}}} \right|}^2}\vartheta _{jlmk}^{'}}}{\left( {1 + {\lambda _{ji_{lm}}}{\vartheta _{jlm}}} \right)^2}\frac{p_{lm} d_j^2\left({\bf z}_{lm} \right) }{M}\xrightarrow[M \to \infty]{a.s.}0.
\end{equation}

Plugging~(\ref{a}),~(\ref{b}) and~(\ref{c}) into~(\ref{temp3}), we have that
\begin{eqnarray}
& & \left| {\hat {\bf h}}_{jjk}^H {\bf \Sigma}_j {\bf  h}_{jlm} \right|^2 \nonumber \\
& \asymp & \frac{p_{jk} d_j^2\left({\bf z}_{jk} \right) }{\left( {1 + {\lambda _{j{i_{jk}}}}{\delta _{jk}}} \right)^2M} \left( {\frac{d_j\left({\bf z}_{lm} \right)}{M}{\rm{tr}} \left({{\bf{T}}}_{jk}^{'}\right) - p_{lm} d_j^2\left({\bf z}_{lm} \right) \lambda_{ji_{lm}} \vartheta_{jlmk}^{'}\vartheta_{jlm} \frac{2+ \lambda_{ji_{lm}} \vartheta_{jlm}}{\left(1+\lambda_{ji_{lm}} \vartheta_{jlm} \right)^2}} \right) \nonumber \\
&=& \frac{p_{jk} d_j^2\left({\bf z}_{jk} \right)d_j\left({\bf z}_{lm} \right)  \mu _{jlmk}}{\left( {1 + {\lambda _{j{i_{jk}}}}{\delta _{jk}}} \right)^2M},
\end{eqnarray}
where $\mu_{jlmk}=\frac{1}{M}{\rm{tr}} ({\bf T}_{jk}^{'}) - p_{lm} d_j({\bf z}_{lm} ) \lambda_{ji_{lm}} \vartheta_{jlmk}^{'}\vartheta_{jlm} \frac{2+ \lambda_{ji_{lm}} \vartheta_{jlm}}{(1+\lambda_{ji_{lm}} \vartheta_{jlm} )^2}$ is defined. Consequently, we have by the dominated convergence theorem that
\begin{equation}
\Eset \left\{\left| {\hat {\bf h}}_{jjk}^H {\bf \Sigma}_j {\bf  h}_{jlm} \right|^2 \bigg | {\hat {\bf h}}_{(j)} \right\} - \frac{p_{jk} d_j^2\left({\bf z}_{jk} \right)d_j\left({\bf z}_{lm} \right)  \mu _{jlmk}}{\left( {1 + {\lambda _{j{i_{jk}}}}{\delta _{jk}}} \right)^2M}\xrightarrow[M \to \infty]{a.s.}0.
\end{equation}
\vspace{-1ex}
\subsection{Noise power}
The noise term in~(\ref{sinr_ul}) is scaled by $\|{\bf g}_{jk} \|^2$ for which we have that
\begin{eqnarray}
\left\|{\bf g}_{jk} \right\|^2 &=& {\hat {\bf h}}_{jjk}^H {\bf \Sigma}_j {\bf \Sigma}_j {\hat{\bf h}}_{jjk} \mathop
= \limits^{\left( a \right)} p_{jk} d_j^2\left({\bf z}_{jk} \right) \frac{\frac{1}{M^2} {\hat {\bf h}}_{{\cal V},ji_{jk}}^H{\bf{\Sigma }}_{jjk}^{'}{\bf{\Sigma }}_{jjk}^{'}{\hat {\bf h}}_{{\cal V},ji_{jk}}}{{\left( {1 + {\lambda _{j{i_{jk}}}}{\hat {\bf h}}_{{\cal V},ji_{jk}}^H{\bf{\Sigma }}_{jjk}{\hat {\bf h}}_{{\cal V},ji_{jk}}} \right)}^2} \nonumber \\
&\mathop  \asymp \limits^{\left( b \right)}& p_{jk} d_j^2\left({\bf z}_{jk} \right) \frac{{{\rm{tr}}\left( {{\tilde{\bf \Phi}}_{{\cal V},ji_{jk}}{\bf{T}}_{jk}^{''}} \right)}}{{{{\left( {1 + {\lambda _{j{i_{jk}}}}{\delta _{jk}}} \right)}^2}{M^2}}} \mathop = \limits^{\left( c \right)} \frac{p_{jk} d_j^2\left({\bf z}_{jk} \right) {\vartheta _{jk}^{''}}}{{{{\left( {1 + {\lambda _{j{i_{jk}}}}{\delta _{jk}}} \right)}^2}M}},
\end{eqnarray}
where step $(a)$ follows from Lemma~\ref{lemma1} and step $( b )$ follows from Lemma~\ref{lemma4} $2 )$, Lemma~\ref{lemma3} and Theorem~\ref{theorem2}. ${{\mathbf{T}}_{jk}^{''}}= {{\mathbf{T}}_{jk}^{''}}(\alpha )$ is given by Theorem~\ref{theorem2} for $\alpha = \frac{\sigma^2 + \varphi_j}{M}$, ${\bf \Theta}={\bf I}_M $, ${\bf D}={\tilde{\bf \Phi}}_{{\cal V},ji_{jk}}$, and ${\bf R}_b =\lambda_{jb}{ \tilde{\bf \Phi}}_{{\cal{V}},jb}$. In step $( c)$, ${\vartheta _{jk}^{''}}=\frac{1}{M}{\text{tr}( {{\tilde{\bf \Phi}}_{{\cal V},ji_{jk}}{\mathbf{T}}_{jk}^{''}})}$ is defined. Then by the dominated convergence theorem, we have
\begin{equation}
\Eset \left\{\left\|{\bf g}_{jk}^H \right\|^2 \bigg | {\hat{\bf h}}_{(j)} \right\} -\frac{p_{jk} d_j^2\left({\bf z}_{jk} \right) {\vartheta _{jk}^{''}}}{{{{\left( {1 + {\lambda _{j{i_{jk}}}}{\delta _{jk}}} \right)}^2}M}} \xrightarrow [M\to \infty]{a.s.} 0.
\end{equation}
Finally, by the continuous mapping theorem, we have
\begin{equation}
{{\eta }_{jk}^{\rm{ul}}} -\frac{\tau_{jk} p_{jk} d_j^2\left({\bf z}_{jk} \right){\delta _{jk}^2}}{{\delta_{jk}^2\sum\limits_{\left( {l,m} \right) \ne \left( {j,k} \right),{i_{_{lm}}} = {i_{jk}}} {\tau_{lm} p_{lm} d_j^2\left({\bf z}_{lm} \right)}  + \sum\limits_{{i_{_{lm}}} \ne {i_{jk}}} { \tau_{lm} d_j\left({\bf z}_{lm} \right)\frac{\mu_{jlmk}}{M}}  + \frac{{{\sigma ^2}}}{M}{\vartheta^{''}_{jk}}}}\xrightarrow [M\to \infty]{a.s.} 0,
\end{equation}
which completes the proof.
\vspace{-2ex}
\section{Proof of Theorem~\ref{theorem4}}\label{sec:proof_thr4}
Except for the channel variance ${\rm{var}} \{{\bf h}_{jjk}^H {\bf w}_{jk}\} =\Eset\{|{\bf h}_{jjk}^H {\bf w}_{jk} -\Eset\{{\bf h}_{jjk}^H {\bf w}_{jk} \} |^2 \}$, large-scale approximations of the signal power and the interference in~(\ref{sinr_dl}) can be calculated by following similar procedures as in Appendix~\ref{sec:proof_thr3}. Thus, only the variance of the effective channel is considered here and we show that it goes to zeros as $M\to \infty$.

Define $c={\hat {\bf h}}_{jjk}^H {\bf \Sigma}_{j}{\hat {\bf h}}_{jjk}$, ${\bar c} = \Eset\{{\hat {\bf h}}_{jjk}^H{\bf \Sigma}_{j}{\hat {\bf h}}_{jjk}\}$, and $b={\tilde{\bf h}}_{jjk}^H {\bf \Sigma}_{j}{\hat {\bf h}}_{jjk}$, then
\begin{eqnarray}
{\rm{var}} \left\{{\bf h}_{jjk}^H {\bf w}_{jk}\right\}& =&\frac{1}{\gamma_{jk}}{\Eset_{\left\{ {\bf{h}} \right\}}}\left\{ \left| {\bf{h}}_{jjk}^H{\bf{g}}_{jk}-{ {\Eset\left\{ {\bf{h}}_{jjk}^H {\bf{g}}_{jk}\right\}}} \right|^2 \right\}  \nonumber \\
 &= &\frac{1}{\gamma_{jk}} \Eset\left\{\left| c - {\bar c} + b \right|^2 \right\} = \frac{1}{\gamma_{jk}}\Eset\left\{\left( c - {\bar c} \right)\left( c + {\bar c} \right) \right\} + \frac{1}{\gamma_{jk}}\Eset\left\{ \left| b\right|^2 \right\},
\end{eqnarray}
where the last step is due to the fact that ${\hat {\bf h}}_{jjk}$ is independent of ${\tilde {\bf h}}_{jjk}$ and that $\Eset \left\{b\right\}=0$.

From step $(a)$ of Eqn.~(\ref{x}), we have %$c \le \theta = \frac{p_{jk} d_j^2\left({\bf z}_{jk} \right) }{\lambda_{ji_{jk}}}$ since
\begin{equation}
c =  \frac{ p_{jk} d_j^2\left({\bf z}_{jk} \right) {\hat {\bf h}}_{{\cal V},ji_{jk}}^H {\bf \Sigma}_{jjk} {\hat {\bf h}}_{{\cal V},ji_{jk}} }{1+\lambda_{ji_{jk}}
{\hat {\bf h}}_{{\cal V},ji_{jk}}^H {\bf \Sigma}_{jjk} {\hat {\bf h}}_{{\cal V},ji_{jk}} } \le \frac{ p_{jk} d_j^2\left({\bf z}_{jk} \right) {\hat {\bf h}}_{{\cal V},ji_{jk}}^H {\bf \Sigma}_{jjk} {\hat {\bf h}}_{{\cal V},ji_{jk}} }{\lambda_{ji_{jk}}
{\hat {\bf h}}_{{\cal V},ji_{jk}}^H {\bf \Sigma}_{jjk} {\hat {\bf h}}_{{\cal V},ji_{jk}} } \le \frac{p_{jk} d_j^2\left({\bf z}_{jk} \right) }{\lambda_{ji_{jk}}} \triangleq \theta.
\end{equation}
Therefore, $c \le \theta$ and same bound also holds for $\bar c$. Thus we have
\begin{equation}
{\rm{var}} \left\{{\bf h}_{jjk}^H {\bf w}_{jk}\right\} \le \frac{2 \theta}{\gamma_{jk}} \Eset\left\{ |c - {\bar c}| \right\} + \frac{1}{\gamma_{jk}}\Eset\left\{ \left| b\right|^2 \right\}.
\end{equation}
It is shown by Lemma~\ref{lemma5} that $c- \frac{d_j^2({\bf z}_{jk} ) \delta_{jk}}{1+\lambda_{ji_{jk}}\delta_{jk}} \xrightarrow[M \to \infty]{a.s.}0$. Since $c$ and $\bar c$ are bounded, this implies by the dominated convergence theorem that $\Eset\{ |c - {\bar c}| \} \to 0$ as $M \to \infty$.

Furthermore, we have that
\begin{eqnarray}
\Eset\left\{\left| b\right|^2 \right\} &= &\Eset\left\{{\hat {\bf h}}_{jjk}^H {\bf \Sigma}_{j}{\tilde{\bf h}}_{jjk} {\tilde {\bf h}}_{jjk}^H{\bf \Sigma}_{j}{\hat {\bf h}}_{jjk} \right\}
 = \Eset \left\{{\hat {\bf h}}_{jjk}^H{\bf \Sigma}_{j} {\bf C}_{jjk} {\bf \Sigma}_{j}{\hat {\bf h}}_{jjk} \right\}\nonumber \\
&\mathop \le \limits^{\left( a \right)}& \frac{1}{\varphi_j^2} \Eset\left\{{\hat {\bf h}}_{jjk}^H{\bf C}_{jjk} {\hat {\bf h}}_{jjk}\right\}
= \frac{1}{\varphi_j^2} {\rm {tr}} \left({\bf \Phi}_{jjk} {\bf C}_{jjk}\right),
\end{eqnarray}
where step $( a )$ holds because ${\bf \Sigma}_j \preceq \frac{1}{\varphi_j}{\bf I}_M$ (where ${\bf A} \preceq {\bf B}$ means that $\bf {B - A}$ is positive semi-definite). Since $\varphi_j^2$ scales as $K^2$ or equivalently as $M^2$, and ${\rm {tr}} ({\bf \Phi}_{jjk} {\bf C}_{jjk})$ scales as $M$, we have that $\Eset\{| b |^2 \} \to 0$ as $M \to \infty$. Consequently,
\begin{equation}
{\rm{var}} \left\{{\bf h}_{jjk}^H {\bf w}_{jk}\right\}\xrightarrow[M \to \infty]{} 0.
\end{equation}
%\vspace{-5ex}
\section{Proof of Theorem~\ref{theorem5}} \label{sec:proof_thr5}
To prove Theorem~\ref{theorem5}, we first rewrite the large-scale approximation of SINR ${\bar \eta}_{jk}^{\rm{ul}}$ in a more tractable way by only having one index instead of two. Define a diagonal matrix ${\bf D} \in \Cset^{LK \times LK}$ and a matrix ${\bf F} \in \Cset ^{LK \times LK}$ as
\begin{align}
&{D}_{k+\left(j-1\right)K,k+\left(j-1\right)K} = { \frac{p_{jk}d_j^2\left({\bf z}_{jk}\right)\delta_{jk}^2}{\vartheta_{jk}^{''}}}, \label{D}\\
%\end{align}
%\begin{align}
&F_{k+\left(j-1\right)K,m+\left(l-1\right)K} = \left\{
\begin{array}{lcl}
0, & {\text{if}}&\,\,\left(l,m\right) = \left(j,k\right),  \\
\frac{\delta_{jk}^2 p_{lm}d_j^2\left({\bf z}_{lm} \right)}{\vartheta_{jk}^{''}}, & {\text{if}}&\,\, i_{lm} = i_{jk}, \left(l,m\right) \ne \left(j,k\right),  \\
\frac{d_j\left({\bf z}_{lm} \right)\mu_{jlmk}}{M\vartheta_{jk}^{''}}, & {\text{if}}&\,\, i_{lm} \ne i_{jk},
\end{array}
\right. \label{F}
\end{align}
respectively, where $[\cdot]_{i,j}$ represents the element in the $i$th row and the $j$th column of the corresponding matrix.

Furthermore, define the vectors ${\bm \tau} =[\tau_{11},\ldots, \tau_{LK} ]^T \in \Rset^{LK \times 1}$ and the scalar $l=k+(j-1)K \in \{1,...,LK\}$, then the uplink SINR approximation in~(\ref{sinr_ul_determ}) can be expressed as
\begin{equation} \label{r}
{\bar \eta}_{jk}^{\rm{ul}} = \frac{\tau_l D_{l,l}}{\left({\bf F}{\bm \tau} \right)_l + \frac{\sigma^2}{M}},
\end{equation}
where $(\cdot)_l$ denotes the $l$th element of the corresponding vector. Furthermore, define the diagonal matrix ${\bf \Psi}^{\rm{ul}} = {\rm{diag}}\{{\bar \eta}_{11}^{\rm{ul}},\ldots,{\bar \eta}_{LK}^{\rm{ul}}\} \in \Cset^{LK \times LK}$ and the vector ${\bf 1}=[1,\ldots,1]^T \in \Rset^{LK \times 1}$, then the above ${\bar \eta}_{jk}^{\rm{ul}}$ can be written in matrix form as
\begin{equation}
{\bf D}{\bm \tau} = {\bf \Psi}^{\rm{ul}}{\bf F}{\bm \tau} + \frac{\sigma^2}{M}{\bf \Psi}^{\rm{ul}}{\bf 1}.
\end{equation}

Using the same notations from~(\ref{D}) and~(\ref{F}), and defining the vector ${\bm \varrho}=[\varrho_{11},\ldots,\varrho_{LK}]^T \in \Rset^{LK \times 1}$, the downlink SINR approximation in~(\ref{sinr_dl_determ}) can also be expressed in matrix form as
\begin{equation}
{\bf D}{\bm \varrho} = {\bf \Psi}^{\rm{dl}}{\bf F}^T{\bm \varrho} + \frac{\sigma^2}{M}{\bf \Psi}^{\rm{dl}}{\bf 1},
\end{equation}
where ${\bf \Psi}^{\rm{dl}} = {\rm{diag}}\{{\bar \eta}_{11}^{\rm{dl}},\ldots,{\bar \eta}_{LK}^{\rm{dl}}\} \in \Cset^{LK \times LK}$ is a diagonal matrix. By setting the downlink SINRs equal to the uplink SINRs as ${\bf \Psi}^{\rm{ul}}={\bf \Psi}^{\rm{dl}}={\bf \Psi}$, then the uplink and downlink powers must satisfy
\begin{align}
{\bm \tau} &= \frac{\sigma^2}{M}\left( {\bf D} - {\bf \Psi}{\bf F}\right)^{-1}{\bf \Psi}{\bf 1},   \label{up_power}\\
{\bm \varrho} &= \frac{\sigma^2}{M}\left( {\bf D} - {\bf \Psi}{\bf F}^T \right)^{-1}{\bf \Psi}{\bf 1},  \label{downlink_power}
\end{align}
if $( {\bf D} - {\bf \Psi}{\bf F})$ is invertible. From the above two equations, it is straightforward to show that ${\bf 1}^T {\bm \varrho} = {\bm \tau}^T {\bf 1}$, which means that the total power is the same. Moreover, after selecting the uplink power ${\bm {\tau}}$ according to some performance metric, we can compute the matrices ${\bf D}$, ${\bf \Psi}$ and ${\bf F}$ and these will make $\left( {\bf D} - {\bf \Psi}{\bf F}\right)$ invertible. We can then obtain the downlink power ${\bm \varrho}$ according to~(\ref{downlink_power}). In this way, the same sum SE as in the uplink is achieved in the downlink.

\bibliographystyle{IEEEtran}
\linespread{1.0}\selectfont
\bibliography{mmse}

\end{document}